\documentclass[useAMS,usenatbib,fleqn]{mn2e}
\usepackage{amsmath,amssymb, aas_macros,times}
\usepackage{subcaption}
\usepackage{graphicx}
\usepackage[usenames,dvipsnames]{xcolor}
\usepackage{float}
\usepackage{verbatim}
\usepackage{url}

\newcommand{\comm}[1]{}
\newcommand{\be}{\begin{equation}}
\newcommand{\ee}{\end{equation}}
\newcommand{\ba}{\begin{eqnarray}}
\newcommand{\ea}{\end{eqnarray}}

\newcommand{\bse}{\begin{subequations}}
\newcommand{\ese}{\end{subequations}}
\newcommand{\bwt}{\begin{widetext}}
\newcommand{\ewt}{\end{widetext}}

\def \eps{\epsilon}

\def \p{\prime}

\DeclareGraphicsExtensions{.eps}

\begin{document}
\title[]{Modelling synchrotron and synchrotron self-Compton emission of gamma-ray burst afterglows from radio to very-high energies}

\author[Joshi and Razzaque]{
Jagdish C. Joshi $^{1,2,3}$ \thanks{jjoshi@nuj.edu.cn} and
Soebur Razzaque $^{1}$ \thanks{srazzaque@uj.ac.za}\\
$^{1}$Centre for Astro-Particle Physics (CAPP) and Department of Physics, University of Johannesburg, PO Box 524, Auckland Park 2006, South Africa\\
$^{2}$School of Astronomy and Space Science, Nanjing University, Nanjing 210093, China\\
$^{3}$Key laboratory of Modern Astronomy and Astrophysics
(Nanjing University), Ministry of Education, Nanjing 210093, China \\
}

\bibliographystyle{mn2e}


\maketitle

\begin{abstract}
Synchrotron radiation from a decelerating blastwave is a widely accepted model of radio to X-ray afterglow emission from gamma-ray bursts (GRBs). GeV gamma-ray emission detected by the Fermi Large Area Telescope (LAT) and the duration of which extends beyond the prompt gamma-ray emission phase, is also compatible with broad features of afterglow emission. We revisit the synchrotron self-Compton (SSC) emission model from a decelerating blastwave to fit multiwavelength data from three bright GRBs, namely GRB~190114C,  GRB~130427A and GRB~090510. We constrain the afterglow model parameters using the simultaneous fit of the spectral energy distributions at different times and light curves at different frequencies for these bursts. We find that a constant density interstellar medium is favored for the short GRB~090510, while a wind-type environment is favored for the long GRB~130427A and GRB~190114C. The sub-TeV component in GRB~190114C detected by MAGIC is the SSC emission in our modelling. Furthermore we find that the SSC emission in the Thomson regime is adequate to fit the spectra and light curves of GRB~190114C.  
For the other two GRBs, lacking sub-TeV detection, the SSC emissions are also modeled in the Thomson regime. 
For the model parameters we have used, the $\gamma\gamma$ attenuation in the blastwave is negligible in the sub-TeV range compared to the redshift-dependent $\gamma\gamma$ attenuation in the extragalactic background light.

\end{abstract}

\maketitle

\section{Introduction}
Afterglow emission occurs in GRBs after the trigger of a burst which produces the prompt emission. The afterglows are important to understand the radiative processes and the source environments in GRBs, located at cosmological distances. The afterglow emission from GRBs were predicted in radio, optical/UV, X-rays and GeV-TeV bands \citep{radio_prediction, 1994ApJ432M, optical_prediction, radiative_xray, panaitescu_meszaros_98, sync_SPN_aftrglw, chiang_dermer_scssc, wind_loss_etc, zhang_meszaros_IC, sari_IC_paper, shape_of_spectral_breaks_G_and_S,  review1,  Kumar_Zhang_15}. The discovery of X-ray and optical afterglow emission from GRB 970228 gave the first hint for the success of GRB afterglow models \citep{costa_xray_97, paradijs_xray_97}. Most of the afterglow radiation features are usually explained using the synchrotron model by \citet{sync_SPN_aftrglw}. More recently the synchrotron models have been successful to interpret Fermi-LAT observations of late GeV emission from GRBs \citep{2009MNRAS.400L..75K, 2010MNRAS.403..926G, 2010OAJ.....3..150R, 2010ApJ714799P}; see also \citet{2013FrPhy...8..661G} for reviews of GeV emission. The recent detection of a sub-TeV spectral component from GRB 190114C and GRB 190829A compliments the expectation of the GRB
afterglow models \citep{2019Natur455M, 2019Natur464A, 2019ATel13052....1D, 2019Natu575459M, 2019Natur448Z}.

Physical processes in addition to the synchrotron radiation are required once the photons detected from the afterglow reached above the maximum synchrotron energy limit. The most efficient process to produce GeV-TeV emission is upscattering of synchrotron photons by the same electrons, known as the synchrotron self-Compton (SSC) or inverse-Compton emission. The intensity of the self-Compton signals from the blastwaves, when they interact with the circumburst medium, was predicted by \cite{IC_prediction}. More detailed calculations were carried out later on by \citet{chiang_dermer_scssc}, \citet{2000ApJ543}, \citet{zhang_meszaros_IC}, and by \cite{sari_IC_paper}. For the detectability of the SSC component in the afterglow, a higher density; greater than $1 ~\rm{cm^{-3}}$; has been estimated by \cite{sari_IC_paper}. The search for this component in GRBs was performed using the Fermi-LAT data and the SSC process was used to explain the delayed GeV component in GRB afterglows \citep{ssc_wang_130427, grb_model_1, wang_ssc_rm}. More recently, High Energy Stereoscopic System (HESS) detected sub-TeV emission from GRB 180720B and GRB 190829A with high significance \citep{2019Natur464A, 2019ATel13052....1D}. These detections have renewed modelling activities of these bursts \citep[see, e.g.,][]{2019ApJ...883..162F, 2019ApJ...879L..26F, 2019ApJ885, 2019ApJ...880L..27D, 2019arXiv191014049Z, 2020A&A636A, 2020arXiv20}.

In this work we revisit the SSC model by \citet{sari_IC_paper} and show its application to the two GeV bright bursts, namely GRB~090510 and GRB~130427A, and to the MAGIC-detected burst GRB~190114C. The model has been presented for the afterglow emission from the forward shock of an adiabatic blastwave decelerating in a constant density or wind-type environment. We constrain the afterglow model parameters using simultaneous fits to the radio to gamma-ray light curves and spectra at different times after the prompt emission.

The outline of this paper is the following. In Section 2, we discuss the dynamics of the blastwave. In Section 3 we discuss the synchrotron emission model and continue with SSC model in Section 4. In Section 5 we discuss absorption of sub-TeV photons
in the blastwave and apply our model to GRBs in section 6. We discuss our results in Section 7 and conclude our work in Section 8. The derivation of synchrotron self-absorption frequency and numerical values of the model parameters for different blastwave evolution scenarios used in modeling are given in the Appendices.

\section{blastwave Modelling}
The GRB event triggers a blastwave, with injected kinetic energy $E_k$, into the surrounding medium which slows down with time \citep{blandford_mckee_soln_76}. For a generic density profile $\propto AR^{-s}$ of the surrounding medium at a distance $R$ from the explosion center, the blastwave energy is given by $E_k = 8\pi AR^{3-s}\Gamma^2 c^2/(17-4s)$, where $\Gamma$ is the Lorentz factor of the shock front \citep{blandford_mckee_soln_76, wind_loss_etc}. We calculate the deceleration time of the blastwave by equating the blastwave energy to $E_k$ and using time $t = (1+z) R/2\Gamma^2 c$ as measured by an observer \citep{blast_wave1} as
\be
t_{\rm dec, i} = \left[
\frac{17E_k(1+z)^3}{64\pi n m_p c^5\Gamma^8}
\right]^{1/3} = 59.6 (1+z) n_0^{-1/3} \Gamma_{2.5}^{-8/3} E_{55}^{1/3} ~\rm s.
\label{t_dec_ism}
\ee
for the case ($s=0$) of interstellar medium of constant gas density $n$ per cubic centimeter. For numerical values, we have used $n=n_0$ cm$^{-3}$, $E_k = 10^{55}E_{55}$ erg and $\Gamma = 10^{2.5}\Gamma_{2.5}$
(with notation $X = 10^n X_n$). For the case ($s=2$) of wind environment typically used for GRB afterglow modeling, 
\be
t_{\rm dec, w} = 
\frac{9E_k(1+z)}{16\pi A m_p c^3\Gamma^4}
= 13.3\, (1+z) A_{\star}^{-1}\Gamma_{2.5}^{-4} E_{55} ~\rm s.
\label{t_dec_wind}
\ee
For numerical values, we have considered the mass-loss rate by the progenitor star is $\dot M_w = 10^{-5} \dot M_{-5} M_{\odot} \rm yr^{-1}$, having a wind velocity
$v_w = 10^{8} v_8$~cm~s$^{-1}$. Therefore $A = \dot M/(4\pi v_w m_p) = 3.02 \times 10^{35} \rm A_{\star} cm^{-1}$,
where $A_{\star} \equiv \dot M_{-5}/v_8$.

For an observer viewing the blastwave along the line of sight to the center, the expansion takes place with the Lorentz factor of the shocked fluid or gas $\Gamma_{\rm g} = \Gamma/\sqrt{2}$, for a strong shock. The blastwave radius evolution with time after the onset of deceleration is given by   
\be
R_{\rm i}(t) = \frac{16 \Gamma_{\rm g, i}^2(t)c t}{1+z}
\label{R_ism}
\ee
and
\be
R_{\rm w}(t) = \frac{8 \Gamma_{\rm g, w}^2(t)c t}{1+z},
\label{R_w}
\ee
respectively for the ISM \citep{1997ApJ...489L..37S} and wind \citep{1998ApJ...493L..31P, 1998MNRAS.298...87D} environments. Subsequently after the deceleration time, the Lorentz factor of the shocked fluid evolves with time as
\be
\Gamma_{\rm g, i}(t) = \frac{\Gamma}{2^{3/4}} \left(\frac{t_{\rm dec, i}}{t} \right)^{3/8} 
\ee
and
\be
\Gamma_{\rm g, w}(t) = \frac{\Gamma}{2^{3/4}} \left(\frac{t_{\rm dec, w}}{t} \right)^{1/4} 
\ee
respectively for the ISM and wind.

\section{SYNCHROTRON EMISSION}
The electrons accelerated at the external shock region radiate away their energy in the amplified magnetic field \citep[see, e.g.,][]{piran_fm, zhang_meszaros04rev}. The magnetic field  takes away a fraction $\epsilon_B$ of the total shock energy, and can be expressed as (all jet-frame quantities are denoted with primes)
\begin{eqnarray}
B_{\rm i}^{\prime}(t) &=& [32 \pi \epsilon_B n_0 m_p c^2]^{1/2}\Gamma_{\rm g}(t) \nonumber \\
B_{\rm w}^{\prime}(t) &=& [32 \pi \epsilon_B A R^{-2} m_p c^2]^{1/2}\Gamma_{\rm g}(t)
\end{eqnarray}
for the ISM and wind cases, respectively. For convenience, we report numerical values of the model parameters for an adiabatic blastwave expansion in these two different scenarios in the Appendix. We discuss shock-accelerated electron spectrum and characteristic breaks therein next.

\subsection{Characteristic electron Lorentz factors}
We consider that the accelerated electrons follow a power-law spectrum which is defined as $N_e(\gamma^\prime_e) = K \gamma_e^{\prime -p}$, with spectral index $p$ and normalization $K = (p-1)n^{\prime} \gamma_{m}^{\prime^{p-1}}$. The power-law electron energy distribution to model the GRB afterglows can have a broad spectral index in the range of 1.4-2.8 as found in a set of GRBs by \citet{param_range1}. The characteristic Lorentz factor of the accelerated electrons at the forward shock for $p>2$ is given by \citep{sync_SPN_aftrglw},
\begin{equation}
\gamma^{\prime}_m (t) = 
\left[\frac{m_p}{m_e} \epsilon_e \frac{p-2}{p-1} \Gamma_{\rm g}\right]; \, p>2 
\end{equation}
The radiation by electrons in the spectrum has two phases of emission called the fast- and slow-cooling. In the fast-cooling, most of the electrons produce the emission efficiently within the dynamic time, while in the slow-cooling, only the high-energy part of the spectrum, above a cooling Lorentz factor $\gamma^\prime_c$, cools efficiently. The electron spectrum defined above will be modified in the fast-cooling regime as
\begin{align}
 N_e(\gamma^\prime_e) & \propto 
\begin{cases}
  \gamma_e^{\prime -2};~~~~~~~~~~ \gamma^\prime_c \le \gamma^\prime_e \le \gamma^\prime_m\\
  \gamma_e^{\prime -p-1}, ~~~~~~~ \gamma^\prime_e >\gamma^\prime_m \,,
  \end{cases}
\end{align}
and in the slow-cooling regime as
\begin{align}
 N_e(\gamma^\prime_e) & \propto 
\begin{cases}
  \gamma_e^{\prime -p};~~~~~~~~~~ \gamma^\prime_m \le \gamma^\prime_e \le \gamma^\prime_c\\
  \gamma_e^{\prime -p-1}, ~~~~~~~ \gamma^\prime_e >\gamma^\prime_c \,.
  \end{cases}
\end{align}

The cooling Lorentz factor ($\gamma^{\prime}_c$), can be estimated by comparing the total cooling time $t^{\prime}_c  = 6 \pi m_e c/ [\sigma_T B{^{\prime}}^2 \gamma^{\prime}_c (1+Y)]$ with the dynamic or expansion time scale $t^{\prime}_{dyn} = t\Gamma_{\rm g}/(1+z)$ as
\begin{equation}
\gamma^{\prime}_c (t) = \left[\frac{6 \pi m_e c^2 (1+z)}{\sigma_T c B{^{\prime}}^2(t) t \Gamma_{\rm g} (1+Y)}\right] \,.
\end{equation}
Here, $\sigma_T$ is the Thomson cross-section and $Y\equiv L_{ssc}/L_{sy}$ is the Comptonization parameter, which is the ratio between the SSC and synchrotron luminosities. In the case of fast-cooling the Y-parameter can be simply expressed as \citep{sari_IC_paper}
\ba
Y ({\rm fast}) = 
\begin{cases} 
\epsilon_e / \epsilon_B \,;  \epsilon_e / \epsilon_B \ll 1\\
\sqrt{\epsilon_e/ \epsilon_B} \,;   \epsilon_e / \epsilon_B \gg 1 \,
\end{cases}
\ea
We investigate the SSC component in GRBs where slow cooling is needed and we explore $Y\gg 1$ scenario for which we define the expression of $Y$ based on \cite{sari_IC_paper}
\ba
&& Y ({\rm slow}) = \sqrt{\epsilon_e/ \epsilon_B} \nonumber \\
&& \times \begin{cases} 
  (t/t_0^{\rm{ssc}})^{(2-p)/[2(4-p)]} \,; & {\rm Adiabatic-ISM} \\
  (t/t_0^{\rm{ssc}})^{(2-p)/(4-p)} \,; & {\rm Adiabatic-Wind}
 
\end{cases}
\ea
The transition time $t_0$ from the fast- to slow-cooling spectra is defined as $\nu_m(t_0) = \nu_c(t_0)$, and in the presence of SSC cooling of electrons
one needs to use the SSC transition time $t_0^{\rm{ssc}}$ or $t_0^{\rm{IC}}$  \citep{sari_IC_paper}. 
The maximum photon energy emitted by synchrotron cooling is proportional to the saturation Lorentz factor ($\gamma^{\prime}_s$). This is calculated by
equating the accelerating time scale $t^{\prime}_{acc} = \phi \gamma^{\prime}_e m_e c/[e B^{\prime}(t)]$, where $\phi^{-1}$ is the acceleration
efficiency for electrons, with the total cooling time $t^{\prime}_c$ defined earlier as,
\begin{equation}
\gamma^{\prime}_s (t) = \left[\frac{6 \pi e}{\phi \sigma_T B^{\prime}(t) (1+Y)}\right]\,.
\end{equation}
Typically $\phi=10$ is assumed and $\phi=1$ correspond to the maximum efficiency. Again, we report numerical values and parameter dependence of the characteristic Lorentz factors for different fireball evolution scenarios in the Appendix.

\subsection{Synchrotron spectra and break frequencies}
The synchrotron break frequencies for the electron Lorentz factors $\gamma_e^{\prime}$ are related by the expression \citep{razzaque_13},
\begin{equation}
 h \nu(t) = \frac{3}{2} \frac{B^{\prime}(t)}{B_{Q}} m_e c^2 \frac{\Gamma_{\rm g}(t)}{1+z} \gamma_e^{\prime^2}
\label{g_fre}
\end{equation}
where $B_Q = 4.41 \times 10^{13}$ G. Using equation~(\ref{g_fre}) we can calculate the synchrotron break frequencies for the minimum ($\nu_m^{\prime}$),
cooling ($\nu_c^{\prime}$) and saturation ($\nu_s^{\prime}$) Lorentz factors $\gamma_m^{\prime}$, $\gamma_c^{\prime}$ and $\gamma_s^{\prime}$,
respectively. These frequencies in the jet frame are transformed to the observer frame by the relations $\nu = \nu^{\prime}\Gamma/(1+z)$.
The synchrotron radiation spectrum from these electrons is distributed in particular frequency order, depending on the fast- and slow-cooling
\citep{sync_SPN_aftrglw, shape_of_spectral_breaks_G_and_S, jessy_moh_razzaque}. The flux of synchrotron radiation is given in the fast-cooling case as  
\begin{align}
 F_{\nu, \rm{fast}} &= f_{\nu, \rm{max}}
\begin{cases}
  (\frac{\nu}{\nu_a})^{2}(\frac{\nu_a}{\nu_c})^{1/3};~~~~~~~~~~~~~\nu < \nu_a\\
  (\frac{\nu}{\nu_c})^{1/3};~~~~~~~~~~~~~~~~~~~~~~~\nu_a\le  \nu \le \nu_c\\
  (\frac{\nu}{\nu_c})^{-1/2};~~~~~~~~~~~~~~~~~~~~~ \nu_c < \nu < \nu_m\\
    (\frac{\nu_m}{\nu_c})^{-1/2} (\frac{\nu}{\nu_m})^{-p/2};~~~~~~~~~ \nu \ge \nu_m \,,
  \end{cases}
  \label{fast_spectrum}
\end{align}  
and in the slow-cooling as
\begin{align}
 F_{\nu, \rm{slow}} &= f_{\nu, \rm{max}}
\begin{cases}
  (\frac{\nu}{\nu_a})^{2}(\frac{\nu_a}{\nu_m})^{1/3};~~~~~~~~~~~~~~\nu < \nu_a\\
  (\frac{\nu}{\nu_m})^{1/3};~~~~~~~~~~~~~~~~~~~~~~~\nu_a \le \nu \le \nu_m\\
  (\frac{\nu}{\nu_m})^{-(p-1)/2};~~~~~~~~~~~~~~ \nu_m < \nu < \nu_c\\
    (\frac{\nu_c}{\nu_m})^{-(p-1)/2} (\frac{\nu}{\nu_c})^{-p/2};~~~~~~ \nu \ge \nu_c \,.
  \end{cases}
  \label{slow_spectrum}
\end{align}
Here $f_{\nu, \rm{max}}$ is the maximum synchrotron flux density which is defined as \citep{sync_SPN_aftrglw, razzaque_13},
\be
f_{\nu,\rm{max}} = \frac{N}{4 \pi d_L^2} \frac{P(\gamma_m^{\prime})}{\nu_m^{\prime}} \frac{\Gamma_{\rm g}^2}{(1+z)^2},
\label{fnu_max_sy}
\ee
with the synchrotron power at $\gamma_{m}^{\prime}$ is given by $P(\gamma_m^{\prime}) = c \sigma_T B^{\prime 2} \gamma{_m^\prime}^2/ 6 \pi$ \citep{radiat_process_Ry_Li}.
The total number of electrons in the blastwave is given by $N=(4/3)\pi R^3 n$, and the luminosity distance to the source is given by $d_L$.
The time-dependence of the synchrotron flux is governed by the time-dependence of $f_{\nu,\rm{max}}$ and of various break frequencies.
Depending on a particular frequency band being observed, the break frequencies can pass through that band at different times.
Two particularly interesting frequencies are $\nu_m$ and $\nu_c$, and the time they appear in the  spectrum $t_m$ and $t_c$, respectively,
are reported in the Appendix for the two different blastwave evolution scenarios. The time and frequency evolution of the flux, denoted as
$F_\nu \propto t^{\alpha}\nu^{\beta}$, give rise to particular relations between $\alpha$ and $\beta$ for different segments in
equations~(\ref{fast_spectrum}) and (\ref{slow_spectrum}). We report these so-called closure relations
\citep{sync_SPN_aftrglw, shape_of_spectral_breaks_G_and_S, ppp_paper} for the synchrotron flux in Table~\ref{Tab:alpha-beta}.
The maximum flux $f_{\nu,\rm{max}}$, synchrotron self-absorption frequency $\nu_a$ and various other break frequencies are reported in the Appendix.

\section{SYNCHROTRON SELF-COMPTON EMISSION}
The SSC spectrum for the same electrons up-scattering synchrotron photons in the Thomson regime is analytically approximated by power-law segments with break frequencies given, following \citet{sari_IC_paper}, by
\ba
\nu_a^{\rm ssc} &\approx& 2\gamma_m^{{\prime}^2}\nu_a \nonumber \\
\nu_m^{\rm ssc} &\approx& 2\gamma_m^{{\prime}^2}\nu_m \nonumber \\
\nu_c^{\rm ssc} &\approx& 2\gamma_c^{{\prime}^2}\nu_c 
\label{IC_freq}
\ea
For this component of the spectrum we follow a similar flux distribution as for the synchrotron part with a shift in frequency as defined above. Similar to the maximum synchrotron flux $f_{\nu,\rm{max}}$ in equation~(\ref{fnu_max_sy}), we define the maximum SSC flux, which is based on the formalism discussed in \cite{zhang_meszaros_IC}, as
\be
f_{\nu,\rm{max}}^{\rm{ssc}} = \frac{\nu_m^{\rm{sy}}}{\nu_m^{\rm{ssc}}} \frac{U^{\prime}_{\rm{ph}}}{U^{\prime}_{\rm{B}}} f_{\nu, \rm{max}}\,.
\ee
Here the magnetic energy density is $U_{\rm B}^{\prime}= B^{\prime^2}/8 \pi$ and the photon energy density is $U_{\rm ph}^{\prime} = (16/3) \sigma_T U_{\rm B}^{\prime} \gamma_m^{\prime 2} R(t)n(R)$.

We also calculate the SSC spectra using the smooth approximation discussed by \cite{sari_IC_paper}, where the target synchrotron photon spectrum is integrated. For details of this approximation we refer to the Appendix A of their paper.

The fast- and slow-cooling SSC spectra in the Thomson regime  follow the same ordering as for the synchrotron spectra. From the flux distribution we can calculate its dependence on the frequency and time, $F_{\nu} \propto t^{\alpha} \nu^{\beta}$, for SSC emission in the two scenarios of blastwave expansions \citep{2000ApJ543}. The temporal and frequency dependence of the fluxes are given in Table~\ref{Table_sy}. The Klein-Nishina effect, however, can become important for SSC emission at very-high energies, which we discuss next.

\subsection{Maximum Energy of Photons in Thomson Scattering Regime}
Klein-Nishina effect in the IC scattering is important for electrons with Lorentz factor above $\gamma^\prime \approx m_ec^2/h\nu^\prime$, for scattering photons of frequency $\nu^\prime$ in the jet frame. This corresponds to a maximum or cutoff SSC photon energy in the Thomson regime as
\begin{equation}
 E^{\rm ssc}_{\gamma,\rm cut} \approx \frac{m_e^2c^4}{h\nu}\frac{\Gamma_{\rm g}^2}{(1+z)^2}.
\label{ICcutoff_ISM}
\end{equation}
Photons above this energy are produced inefficiently in the Klein-Nishina regime, where $\nu^{\rm ssc} \approx \gamma^\prime \nu$, and the SSC flux decreases. 
Slow cooling case is described here for which 
the peak synchrotron flux is at $\nu_c$. The corresponding cutoff photon energy via Thomson scattering is 
\begin{eqnarray}
\begin{split}
  E^{\rm ssc}_{\gamma,\rm cut} = 1.1\, (1+z)^{-3/4} n_{0,-5}^{3/4} E_{55}^{3/4} \epsilon_{B,-1}^{3/2} t_2^{-1/4} (1+Y)^2 ~{\rm TeV}
\end{split}
\label{ismkn}
\end{eqnarray}
for a constant-density environment with density $n_0 = 10^{-5}n_{0,-5}$~cm$^{-3}$ and
\begin{eqnarray}
\begin{split}
E^{\rm ssc}_{\gamma,\rm cut} = 5.5 A_{\star,-2}^{3/2} \epsilon_{B,-1}^{3/2}
 t_2^{-1} (1+Y)^2 ~{\rm TeV}
\end{split}
\label{windkn}
\end{eqnarray}
for the wind environment with the wind parameter $A_* = 10^{-2}A_{*,-2}$. Therefore, the Klein-Nishina effect becomes important in the TeV energy range. We explicitly calculate this energy for GRB~190114C later on, indicating that the sub-TeV MAGIC detection can be modeled as SSC emission in the Thomson regime.

Note that these cutoff energies are larger than the photon energies, few hundred GeV, at which absorption due to $\gamma\gamma \rightarrow e^{\pm}$ interactions with the extragalactic background light (EBL) becomes important for cosmological distances \citep{2009ApJ...697..483R, finke_razz_der_09}.Therefore, direct detection of a Klein-Nishina effect in the TeV energy range of the GRB spectra can be difficult.


\section{INTERNAL ABSORPTION IN THE BLASTWAVE}
In this section we discuss absorption of gamma-rays within the forward shock due to $\gamma\gamma\to e^\pm$ interactions with synchrotron photons. The comoving number density of the synchrotron photons with frequency $\nu$ can be calculated from the corresponding observed flux $F_\nu$ as
\be
n_\nu^\prime = \left( \frac{d_L}{R} \right)^2 
\frac{1+z}{\Gamma hc} F_\nu
\ee
As such, the $\gamma\gamma$ optical depth corresponding to that frequency can be calculated in delta-function approximation as
\be
\tau_{\gamma\gamma} = \left( \frac{\sigma_T}{5} \right) \frac{n^\prime_\nu R}{\Gamma_{\rm g}}
\ee
This affects the photons of energy
\be
E_\gamma = \frac{2m^2_ec^4\Gamma_{\rm g}^2}{(1+z)^2h\nu}    
\ee
For an estimate, we use $F_\nu = f_{\nu,{\rm max}}$ at $h\nu_c$ for the slow-cooling spectra to approximate the maximum optical depth for gamma-rays. These give the opacities in the ISM and wind environments as
\begin{eqnarray}
 \tau_{\gamma\gamma, {\rm ISM}} &=& 0.08\,(1+z)^{-1/2} n_{0,-5} \epsilon_{B,-1}^{1/2} E_{55}^{1/2} t_2^{1/2} \\
 \tau_{\gamma\gamma, {\rm wind}} &=& 0.4\, (1+z)^{1/2} A_{*, -2}^2 \epsilon_{B,-1}^{1/2} E_{55}^{-1/2} t_2^{-1/2} 
\label{internalgg}
\end{eqnarray}
respectively. The corresponding gamma-ray energies are
$E_{\gamma} \approx 2E_{\gamma,\rm cut}^{\rm ssc}$, where $E_{\gamma,\rm cut}^{\rm ssc}$ is found from equations (\ref{ismkn}) and (\ref{windkn}), respectively, for the ISM and wind.

We have also calculated the $\gamma\gamma$ optical depth, for the full target photon distribution using \citep{pair_prod_gs}
\begin{eqnarray}
 \tau_{\gamma\gamma}(E_{\gamma}) &=& 
 \frac{R}{\Gamma_{\rm g}} \pi r_0^2  \left[\frac{m_e c^4 \Gamma}{(1+z) E_{\gamma}} \right]^2 \\ \nonumber
&& \times \int_{\frac{m_e c^4 \Gamma_{\rm g}}{(1+z) E_{\gamma}}}^{(1+z)E_{\gamma}/\Gamma_{\rm g}}
 \frac{n^{\prime}(\epsilon^{\prime})}{\epsilon{^{\prime}}^2} \phi[S_0(\epsilon^{\prime})] d\epsilon^{\prime},
\label{fullgg}
\end{eqnarray}
%
%
where $\epsilon^{\prime} = h \nu^{\prime}$ 
and $n^\prime({\epsilon^{\prime}})d\epsilon^\prime = n^\prime_\nu$. The function $\phi[S_0(\epsilon^{\prime})]$ is given in \citet{brown1973ApL203B} and the argument is defined as $S_0(\epsilon^{\prime}) = (1+z) \epsilon^{\prime} E_{\gamma}/\Gamma_{\rm g} m_e^2 c^4$. This calculation assumes that the target photons are distributed isotropically in the blaswave.

\begin{figure}
    \centering
    \includegraphics[width=0.5\textwidth]{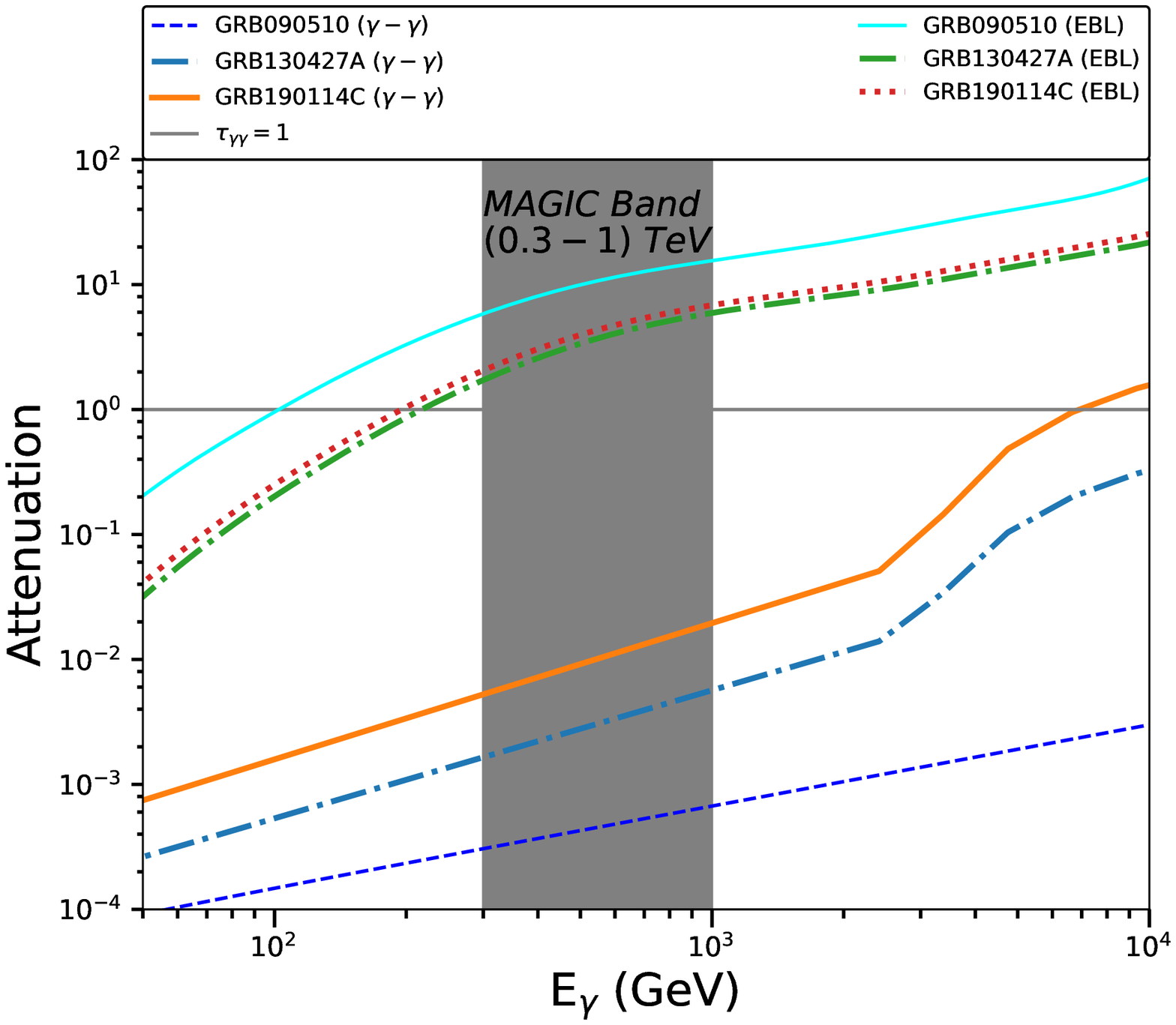}
    \caption{The $\gamma\gamma\to e^\pm$ optical depths in the blastwave and in the EBL.
    The blastwave opacities are calculated for the observed GeV-TeV radiation in the target photon field. These values are estimated at 68~s
    for GRB~190114C, at 352~s for GRB~130427A, and at 100~s for GRB~090510 where the density of target photons is the maximum for each case.
    The EBL optical depths for the model by \citet{finke_razz_der_09}.
}
 \label{opacity}
\end{figure}

Our results from numerical calculations are shown in Figure~\ref{opacity} where we consider the target photon distribution at times $t =100$s for GRB 090510, $t = 352-403$s for GRB 130427A and $t = 68-110$s for GRB~190114C. These values are similar to our analytical calculation of the optical depths, which serve as cross checks. As shown, the internal $\gamma\gamma$ opacity is negligible below the TeV energy range for the set of parameters we have used in this work. As such we do not consider secondary cascade emission.
We have also plotted in Figure~\ref{opacity} the $\gamma\gamma$ opacity due to EBL using \citet{finke_razz_der_09} model. EBL attenuation is significant for $\gtrsim 100$~GeV range and determines the maximum observable photon energy from GRBs studied here.




%
%

\section{Modelling of broad-band afterglow emission}
We investigate the afterglow emission of GeV-bright bursts, namely the short GRB 090510, long GRB 130427A, as well as the recently-detected long GRB 190114C. We describe the details of their emission in next subsections. The parameters used in modelling the afterglow emission of these GRBs, are given in Table \ref{Table_pr}.  We also show the CTA sensitivity in the SEDs and light curves. In the light curves the sensitivity is plotted at 25~GeV and at 250~GeV. In the SEDs, the CTA sensitivity is shown for a duration of 300-1000~s. In such calculations, a $5\sigma$ significance is required in each energy bin and the source flux needs to be few times higher than the background signal. These sensitivities for 25 events in each bin, where 4 bins are taken per decade of energy, is calculated by \citet{cta_sens1}. These calculations show that the differential flux sensitivity of CTA for 25~GeV gamma-rays is approximately $10^{-9}$~erg~cm$^{-2}$~s$^{-1}$ if the transient source lifetime is considered to be within 10 s. In our work we have used the CTA sensitivity for transients calculated at an elevation angle $70^{\circ}$, which has been retrieved from the CTA website\footnote{\url{https://www.cta-observatory.org/science/cta-performance/}}.

\begin{table}
\caption{The afterglow model parameters from simultaneous interpretation of the SED and light curves. 
\label{Table_pr}} 
\begin{tabular}{||cccc||}
\hline
 Parameter &  GRB 090510 & GRB 130427A & GRB 190114C  \\
\hline
\hline
$E_{k} (\rm{erg}) $  & $9 \times 10^{51}$ & $3 \times 10^{54}$&  $4 \times 10^{54}$ \\
$\Gamma$   & 1500 & 280 &  300  \\
$t_{\rm{dec}} (\rm{s})$    & 4.5 & 98.1 &  52.6 \\
$t_0^{\rm ssc} (\rm{s})$     & 0.0001  & 25.5 &  8.6   \\
$t_{\rm {jet}} (\rm{s})$   & 3000  & - &   -  \\
$A_{\star}(\rm{cm^{-1}})$        & - & $1 \times 10^{-2}$ &  $2 \times 10^{-2}$ ($6\times 10^{-2}$)    \\

$n_0 (\rm{cm^{-3}})$     & $1 \times 10^{-5}$ & - &  -    \\

$p$    & 2.3 & 2.05 &  2.18 ($2.1$)  \\
$\epsilon_e$      & 0.2 & 0.27 &  0.033 ($0.034$)  \\
$\epsilon_B$        & 0.02 & $0.015$ &  0.012  ($0.0039$)   \\
$Y({\rm slow})$ & $0.93t_2^{-0.09}$ & $4.1t_2^{-0.03}$ & $1.3t_2^{-0.1}$ \\
$\phi$      &  10     &   1 & 1  \\

\hline
\end{tabular}
\label{Tab:two_comp_jet}
\end{table}

\subsection{Short GRB~090510}
The GRB~090510 with a duration of $T_{90} = 0.3 \pm 0.07$ s was observed in the early afterglow phase by the Swift and Fermi satellites~\citep{swift_fermi_090510}. The redshift of the burst is $z = 0.903 \pm 0.003$ \citep{rs_grb_090510} and the corresponding luminosity distance is $1.8 \times 10^{28}$ cm. These observations were modeled using typical synchrotron radiation \citep{swift_fermi_090510, 2010A&A510L7G, 2011ApJ_wang, 2016ApJ83122F}. A combined electron-proton synchrotron model was used by~\citet{razzaque_aftrg_prot_sync}, where the proton component was used to interpret the Fermi-LAT data. A two-component jet model was used by \citet{corsi_two_bw} to interpret the same data.

\begin{figure}
    \centering
    \begin{subfigure}[t]{0.45\textwidth}
        \centering
        \includegraphics[width=1\linewidth]{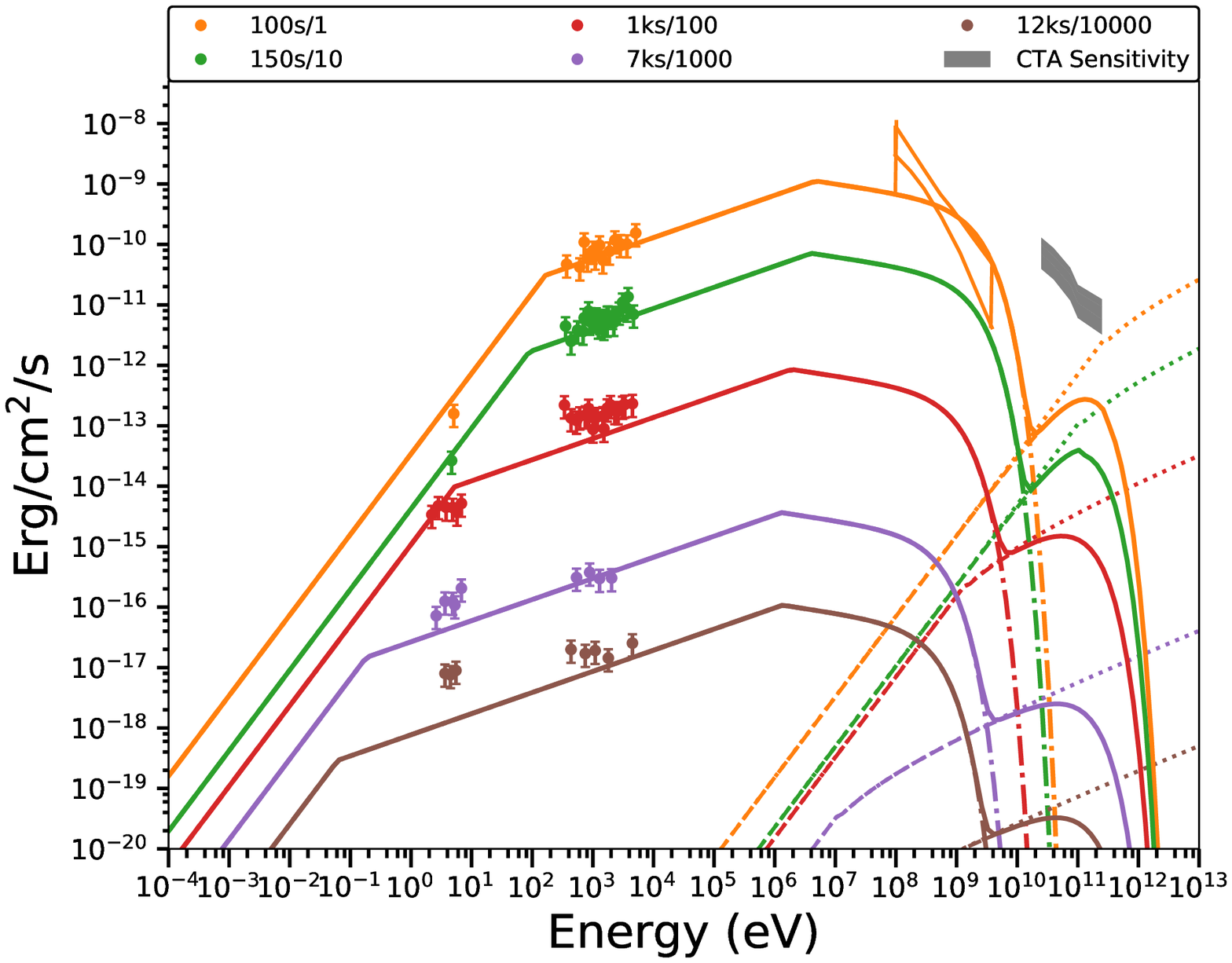}
        \label{sed_grb_090510}
    \end{subfigure}
    \hfill
    \begin{subfigure}[t]{0.45\textwidth}
        \centering
        \includegraphics[width=1\linewidth]{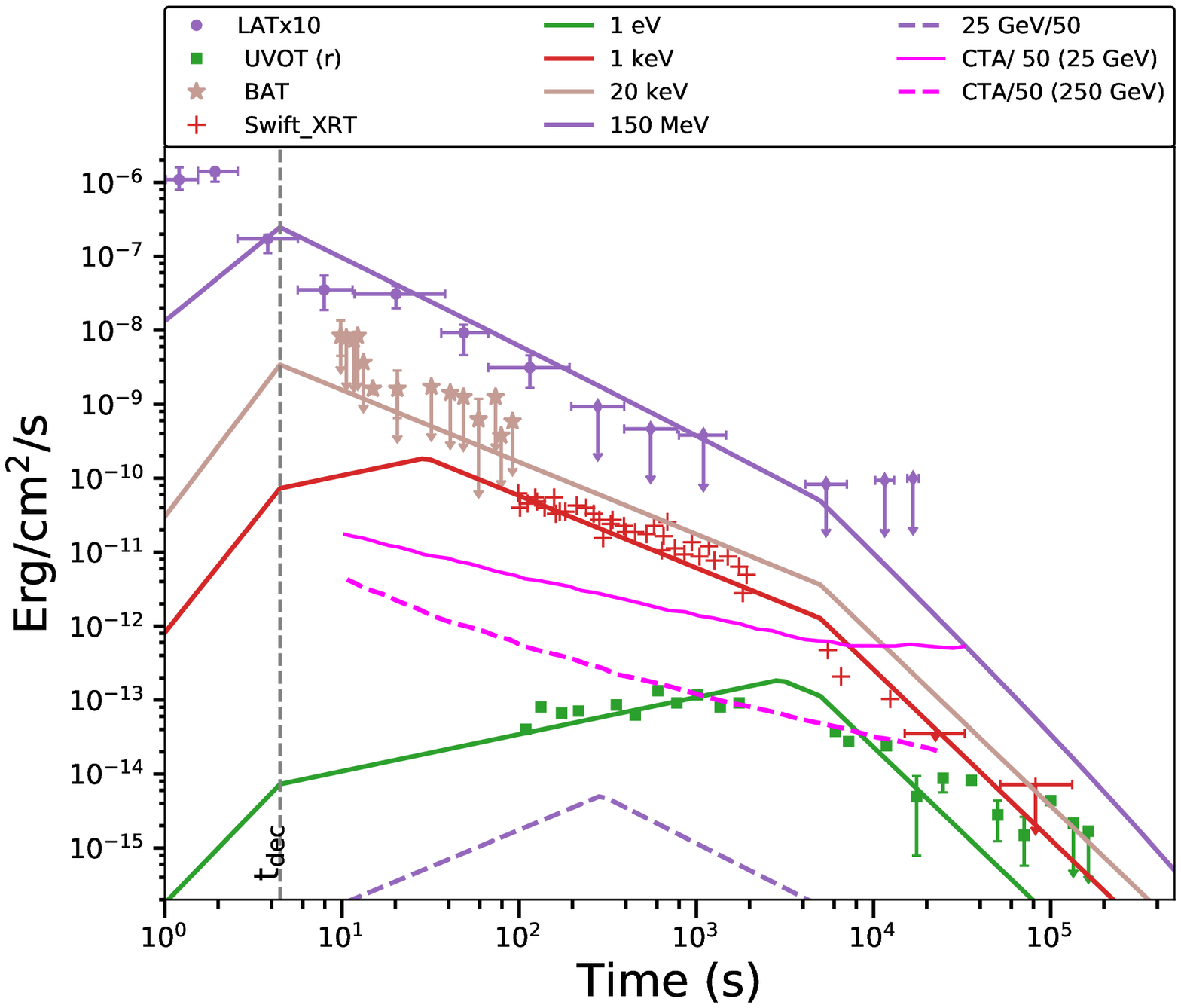}
    \end{subfigure}
    \caption{{\it Top Panel:} The SED of short GRB~090510, where the multiwavelength data are shown for the Swift and Fermi-LAT observations \citep{swift_fermi_090510}.  {\it Bottom Panel:} The Swift/BAT (15-350 keV), Swift/UVOT, and Swift/XRT (0.3-10keV), Fermi-LAT (100 MeV-4 GeV) light curves are shown. For the duration of 1.9-5.1 ks, there are no data points in Swift/XRT observation due to Earth occulation. The data points are taken from \citet{swift_fermi_090510}, SWIFT-XRT database {\it https://www.swift.ac.uk/analysis/xrt/}.
    The SED fluxes are scaled by factors 1, 10,
    $10^2$, $10^3$ and $10^4$ in decreasing order of time.}
    \label{fig_sgrb}
\end{figure}

Figure~\ref{fig_sgrb} shows the data and our model curves for GRB 090510. Our interpretation favours a constant circumburst
environment with a very low  density of $1 \times 10^{-5}$~cm$^{-3}$. The modelling of this source requires slow cooling of the relativistic electrons.
The parameters are shown in Table~\ref{Tab:two_comp_jet} and fast to slow cooling transition occurs at $t_0^{\rm IC} =0.0001$ s. The no jet break model
has surplus of flux in late times, which is corrected using the jet-break feature in this source.
In the jet-break, the time dependence of the break frequencies are $\nu_a \propto t^{-1/5}$, $\nu_m \propto t^{-2}$, and $\nu_c \propto t^{0}$ and the
maximum flux goes as $f_{\nu,max} \propto t^{-1}$. During the jet-break phase the closure relations for slow cooling are
$F_{\nu} \propto \nu^2 t^1$ for $\nu < \nu_a$ and $F_{\nu} \propto \nu^{1/3} t^{2/3}$ for the regime $\nu_a  < \nu < \nu_m $. The late time emission
needs a steeper dependence on time and the emission is explained using the regimes $\nu_m  < \nu < \nu_c $ where $F_{\nu} \propto \nu^{-(p-1)/2} t^{1-p}$
and $\nu > \nu_c$ where $F_{\nu} \propto \nu^{-p/2} t^{1-p}$.
The jet-break time is estimated using $t_{\rm{jet}} = 5 \times 10^{5} (1+z) (E_{55}/n)^{1/3} \theta_{-1}^{8/3}$~s \citep{jetbreak_99_sph}.
We find that the jet-break time is 3000 s, which is in the range of 1.4-5.1 ks, as discussed in \citet{razzaque_aftrg_prot_sync}.


Our modelling confirms the need for very low density ISM medium, as also shown in earlier results by~\citet{corsi_two_bw}. The maximum photon energy due to synchrotron emission with our model parameters is 1.8~GeV at 100~s (see the SEDs plotted in Fig.~\ref{fig_sgrb}). The model for early two epochs i.e. 100 s and 150 s produces slightly higher amount of optical flux. In the LAT energy range and the remaining part of emission is well produced. For the redshift of GRB~090510 the EBL attenuation energy is $\sim 100$ GeV based on the EBL model by \cite{finke_razz_der_09} for which we have also plotted the $\gamma\gamma$ opacity in Fig.~\ref{opacity}. The suppression of the SSC component plotted in Fig.~\ref{fig_sgrb} using smooth approximation is due to the EBL attenuation and $\gamma\gamma$ effects are negligible in the blastwave, based on our input model parameters.

The breaks in the light curves for 1 eV occurs at $t_m \sim 2500$ s and for 1 keV at $t_m \sim 25$ s. We have also shown in the light curve, bottom panel of
Figure \ref{fig_sgrb} the rising part before the deceleration time.  For slow cooling, which is valid for this case, the rising part is defined as
$F_{\nu, s} \propto t^2$ for $\nu < \nu_{a,s}$,  $F_{\nu, s} \propto t^3$ for $\nu_{a,s}  < \nu < \nu_{m,s} $ , $F_{\nu,s} \propto t^3$ for
$\nu_{m,s}  < \nu < \nu_{c,s} $ and $F_{\nu,s} \propto t^2$ for $\nu > \nu_{c,s} $ \citep{1999ApJ520641S, 2013NewAR57141G}. The SSC emission has
the temporal dependence for pre-deceleration
is $F_{\nu, ssc} \propto t^3$ for $\nu_{m, ssc}  < \nu < \nu_{c, ssc} $
and $F_{\nu, ssc} \propto t$ for $\nu > \nu_{c, ssc}$. For optical, XRT and BAT energy range we have $F_{\nu} \propto t^3$ and
for the 100 MeV synchrotron flux it is proportionl to $t^2$, while for SSC emission at 25 GeV, $F_{\nu, ssc} \propto t^3$.

\begin{figure}
    \centering
    \begin{subfigure}[t]{0.45\textwidth}
        \centering
                \includegraphics[width=1\linewidth]{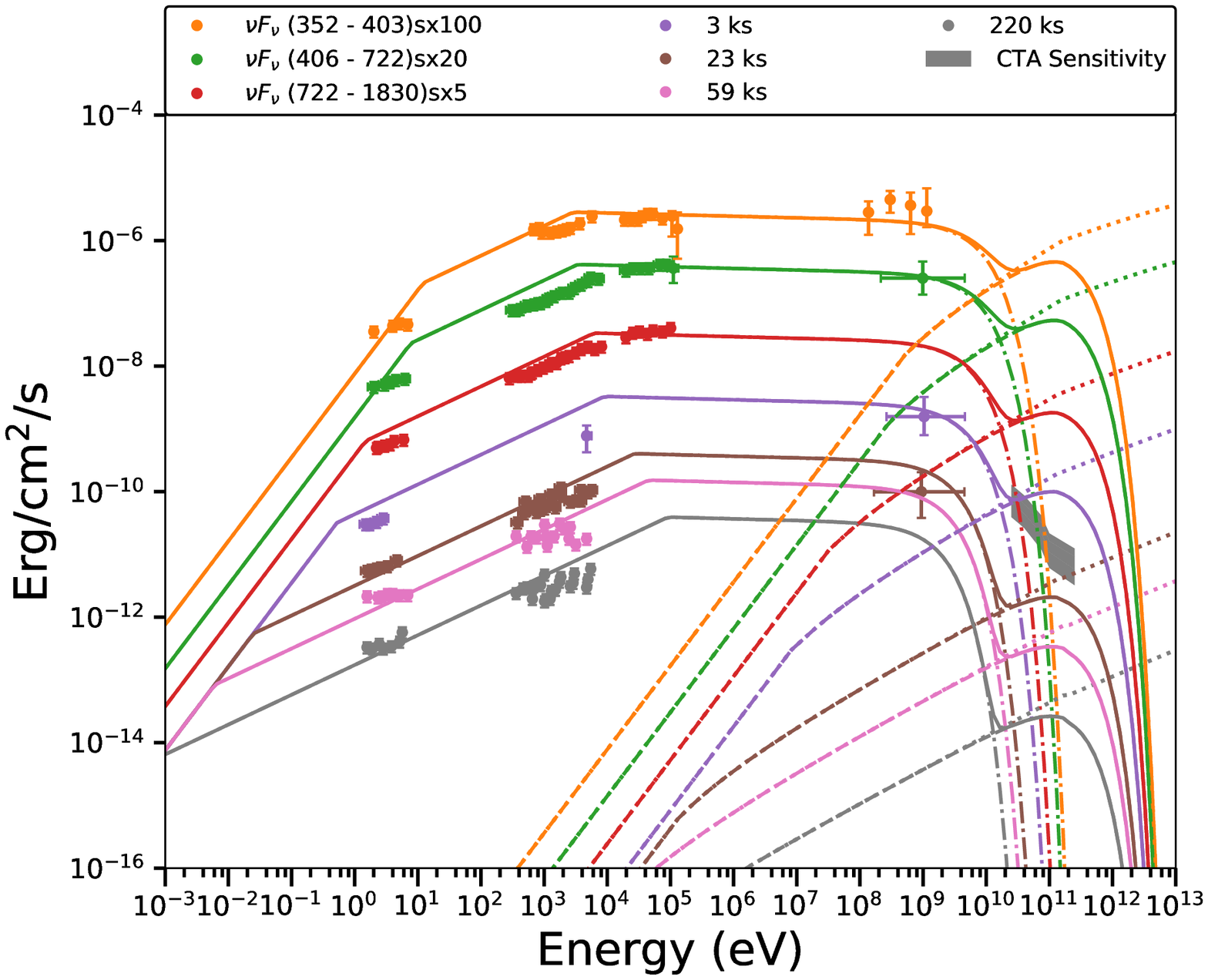}
    \end{subfigure}
    \hfill
    \begin{subfigure}[t]{0.45\textwidth}
        \centering
                \includegraphics[width=1\linewidth]{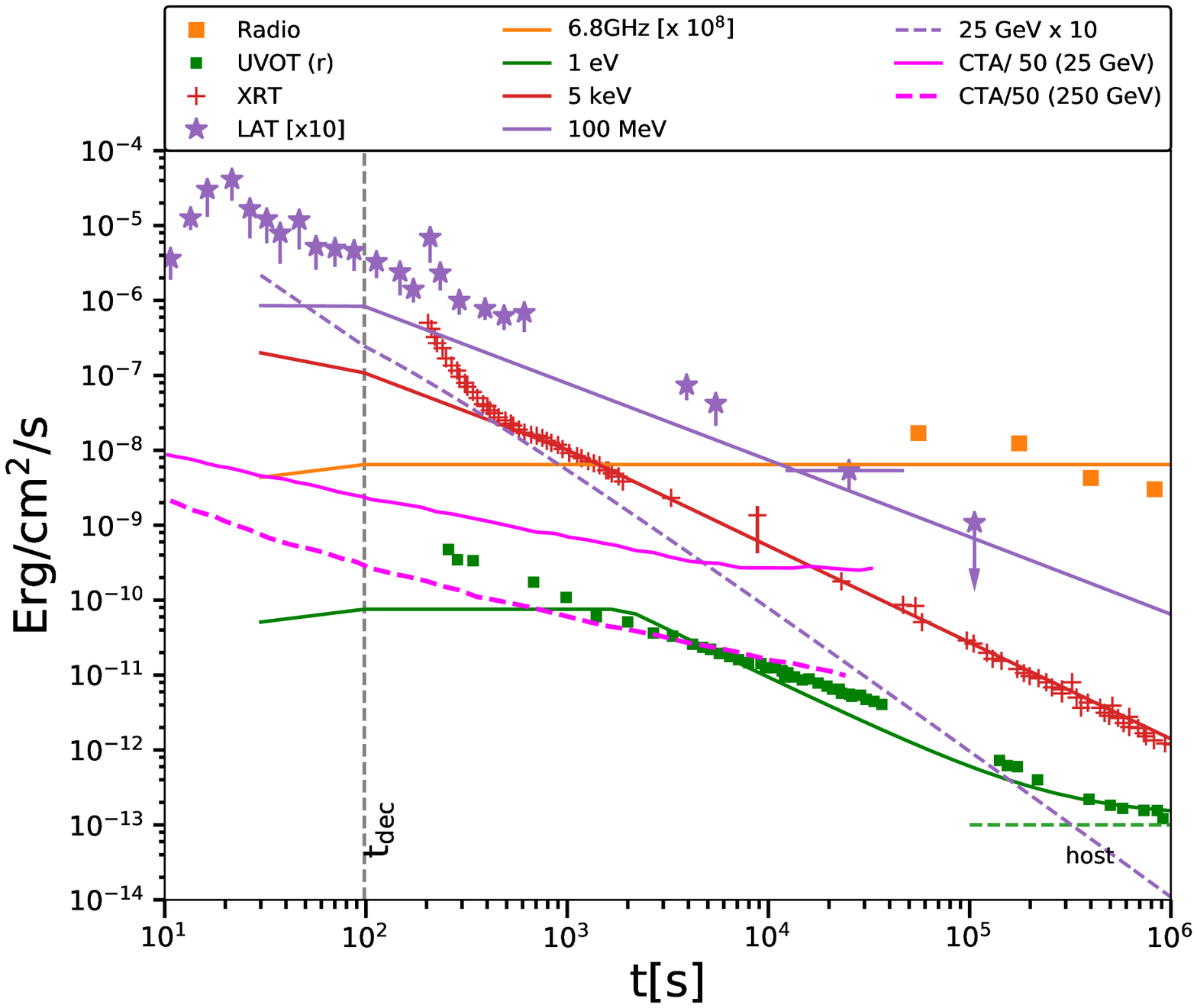} 
    \end{subfigure}
    \caption{{\it Top Panel:} The SED of the long GRB~130427A, where in the optical the total contribution of the forward shock and host galaxy is shown. 
    {\it Bottom Panel:} The light curves in optical to gamma-rays for GRB~130427A.
    The data used in these two plots: Radio 6.8 GHz, UVOT, (0.3-10) keV and LAT (0.1-100) GeV data are taken from \citet{afterglow_maseli} and from SWIFT-XRT database {\it https://www.swift.ac.uk/analysis/xrt/}.}
     \label{grb_130427A_wind}
\end{figure}

\subsection{Long GRB~130427A}
One of the brightest long GRB~130427A with $T_{90} = 276 \pm 5$~s was located at redshift $z=0.34$ \citep{rs_130427a}. The afterglow of GRB~130427A was observed up to 220 ks in radio and optical wavelengths while the X-ray and gamma-ray observations by Swift-XRT and Fermi-LAT were active upto 1.8~ks \citep{afterglow_maseli}. A photon of energy 95 GeV was detected at $T_0 + 244$ s and a $32$ GeV photon was detected in late time at $T_0 + 34.4$ ks \citep{GRB_130427A}. Its association with a type-Ic supernova \citep{typeic} provides us further evidence that long GRB 130427A is produced by the collapse of a massive star. The light curves for this source has been modelled for constant density medium~\citep{grb_model_1, afterglow_maseli, 2013ApJ77695F, ssc_wang_130427, 2013ApJ771L13T} as well as for wind medium~\citep{2013ApJ779L1K, grb_model_1, 2016ApJ818190F}. The reverse shock emission features are also used in some models for this burst \citep{2013ApJ77619L, 2016ApJ818190F,  2014Sci34338V, 2013ApJ776119L, 2014Sci34338V}.


The SED and light curves from our modelling for this burst is shown in Fig.~\ref{grb_130427A_wind}. The maximum photon energy due to synchrotron emission with our model parameters is 7.5~GeV at 352~s. We used a wind environment to explain the multiwavelength observations. The parameters of our model are reported in Table~\ref{Tab:two_comp_jet}. We estimate $E_{\gamma,\rm cut}^{\rm ssc}$ defined in equation (\ref{windkn}) which is $\sim $ 2.2 TeV at 352 s. The EBL attenuation for this redshift is $\sim 300$ GeV and we use Thomson scattering regime in our model.
The early intervals of SED with duration 352-403~s, 403-722~s and 722-1830~s are modelled using times at 352~s, 722~s and 1830~s, respectively.
The later SEDs are plotted with mentioned time in the figure legend. The internal $\gamma\gamma$ opacity is negligible (see Fig.~\ref{opacity}).
Therefore, we have used a cutoff energy of 300~GeV for the SED at all time in Fig.~\ref{grb_130427A_wind}. The breaks in the light curves, for 2~eV
is at $t_m \sim 1352$~s.

The pre-deceleration phase in our light curve, bottom panel of Figure \ref{grb_130427A_wind}
follows the dependence $F_{\nu, s} \propto t^2$ for $\nu < \nu_{a,s}$,  $F_{\nu, s} \propto t^{1/3}$ for $\nu_{a,s}  < \nu < \nu_{m,s} $ , $F_{\nu,s} \propto t^{(1-p)/2}$ for
$\nu_{m,s}  < \nu < \nu_{c,s} $ and $F_{\nu,s} \propto t^{(2-p)/2}$ for $\nu > \nu_{c,s} $ \citep{1999ApJ520641S, 2013NewAR57141G}.
The SSC emission has the temporal dependence for pre-deceleration is $F_{\nu,ssc} \propto t^{(1-p)/2}$ for $\nu_{m,ssc} < \nu < \nu_{c,ssc} $
and $F_{\nu,ssc} \propto t^{3/2}$ for $\nu > \nu_{c,ssc} $. For GRB 130427A before deceleration time we have $t^{1/3}$ dependence
for 6.8 GHz and 2 eV, and $t^{-0.52}$ for 5 keV and $t^{-0.02}$ for 100 MeV light curve. For the SSC light curve at 25 GeV the dependence is
$t^{-0.52}$.

\subsection{Long GRB~190114C}
The sub-TeV GRB~190114C is located at a redshift $z = 0.4245 \pm 0.0005$ \citep{2019GCN237081C}. This is the first case of afterglow observation where a sub-TeV component was observed by the MAGIC ground-based Cherenkov telescope~\citep{2019Natur455M, 2019Natu575459M}. The isotropic gamma-ray energy released in this burst was $(2.5 \pm 0.1) \times 10^{53}$~erg~\citep{2019Natu575459M} and the burst duration is $T_{90} = 116.4 \pm 2.6 $ s for 50-300 keV range \citep{2020ApJ9A}.  The optical light curve in the early afterglow phase has a steeper index and shows the signatures of reverse shock emission \citep{laskar1926L}. It is widely believed that the observed sub-TeV component is the SSC emission from the blastwave.

\begin{figure}
    \centering
    \begin{subfigure}[t]{0.45\textwidth}
        \centering
                \includegraphics[width=1\linewidth]{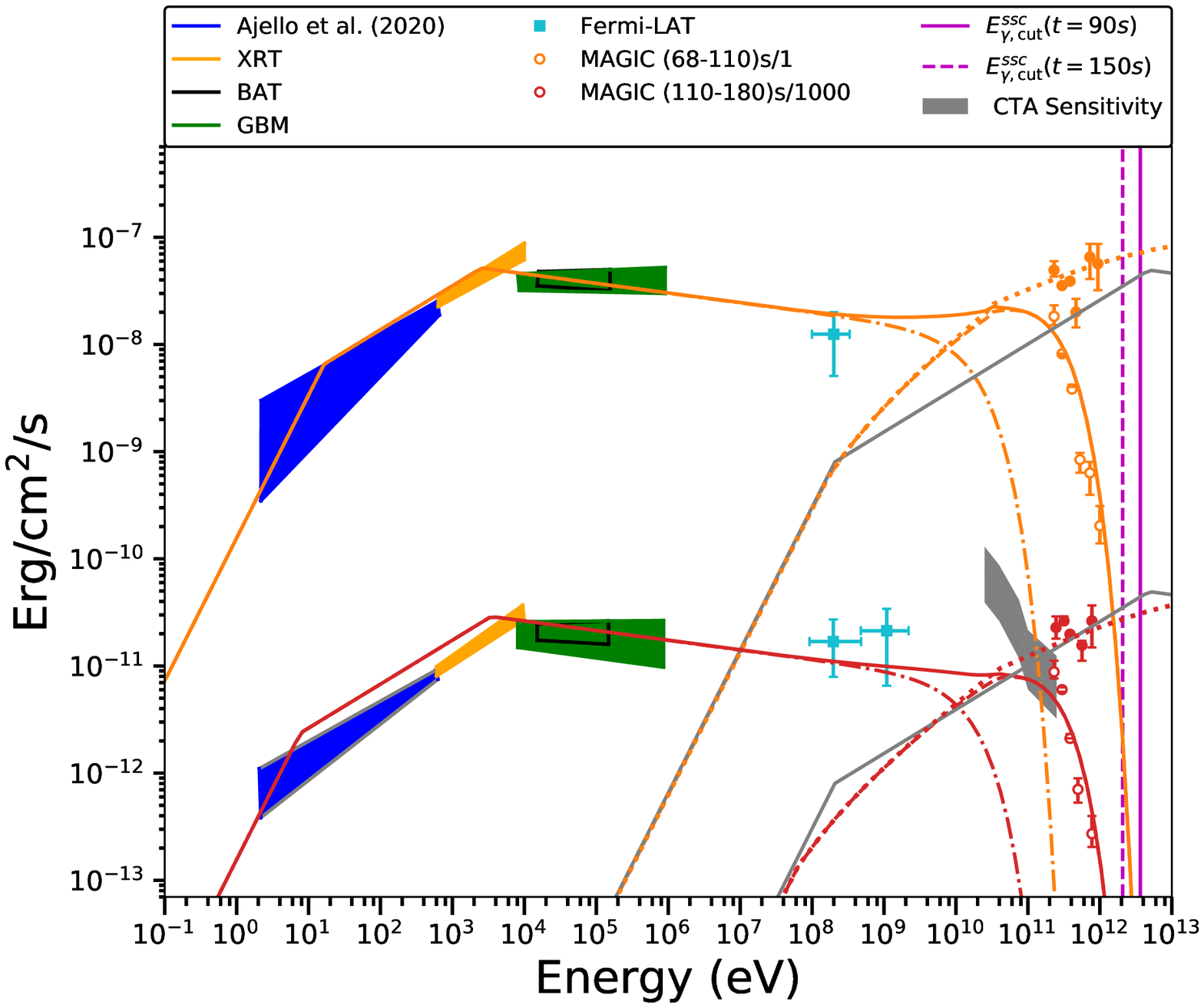}
    \end{subfigure}
    \hfill
    \begin{subfigure}[t]{0.45\textwidth}
        \centering
                \includegraphics[width=1\linewidth]{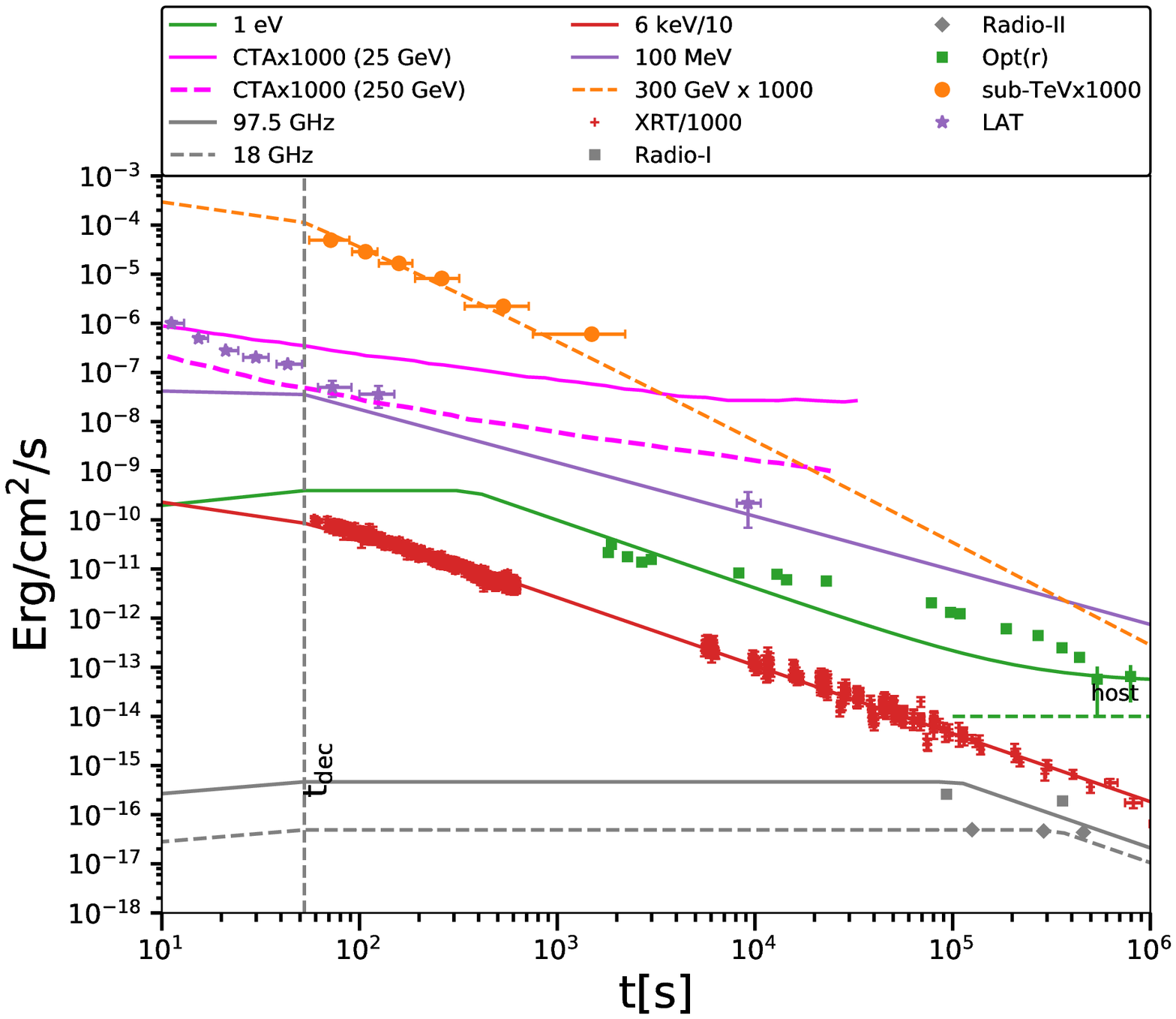} 
    \end{subfigure}
    \caption{{\it Top Panel:} The SED of the long GRB 190114C and we have shown our model fit to MAGIC sub TeV data.
    {\it Bottom Panel:} The data used in these two plots: Radio 18 and 97.5 GHz, XRT (0.3-10) keV, BAT (15-150) keV, GBM (10 - $10^4$) keV, LAT (0.1-100) GeV are
    taken from \citet{2020ApJ9A,2019Natu575459M} and SWIFT-XRT database {\it https://www.swift.ac.uk/analysis/xrt/}. Note that the optical range data is actually a model based extrapolation as described by \citet{2020ApJ9A}.The vertical solid and dashed magenta color lines shows the crossover from Thomson to Klein-Nishina regime.}
     \label{grb_190114_wind}
\end{figure}

We have modelled the SEDs and lightcurves of GRB~190114C using an adiabatic blastwave in a wind environment. The SED and light curves from our modelling for this burst is shown in Fig.~\ref{grb_190114_wind}. 

The SSC spectra are shown using analytical approximation (grey solid lines) and also using the smooth approximation (dotted orange curves). The MAGIC data in the SED with empty circles are the observed ones while the filled circles are the ones corrected for the EBL attenuation. In the light curves the MAGIC data are corrected for the EBL and the corresponding model output is plotted using smooth approximation for SSC emission. 
The sub-TeV components for the intervals 68-110~s and 110-180~s are modelled using times at 90~s and 150~s respectively. The two vertical lines correspond to the cutoff energy defined in equation (23) for these two intervals, i.e. $ E_{\gamma, \rm cut}^{\rm ssc}(t =90 \rm s) \sim 3.7 $ TeV and $ E_{\gamma, \rm cut}^{\rm ssc}(t =150 \rm s) \sim 2.1 $ TeV. This indicates that the Klein-Nishina effect for the SSC emission becomes important at energies above few 100 GeV, for some combination of afterglow model parameters. Hence, we can model the MAGIC detected photons from GRB 190114C in the Thomson scattering regime. It can also be seen from Fig.~\ref{opacity} that the internal $\gamma\gamma$ opacity is negligible and very high-energy photons are attenuated in the EBL. To model the SEDs we have used an EBL cutoff energy of 200~GeV at which the EBL opacity is $\sim 1$ for the model by~\citet{finke_razz_der_09}. 
In the light curve for this source in Fig.~\ref{grb_190114_wind}, only in optical bands the flux becomes harder for times $10^4 - 5 \times 10^5$~s and it cannot be explained using our one-zone model. The breaks in the light curves, for 97.5~GHz is at $t_m \sim 10^4$~s while for 18 GHz at $t_a = 732 $~s and $t_m \sim 3 \times 10^4$~s.
For GRB 190114C before deceleration time we have $t^{1/3}$ dependence
for 18 and 97.5 GHz and 1 eV, and $t^{-0.6}$ for 5 keV and $t^{-0.1}$ for 100 MeV light curve. For the SSC light curve at 300 GeV the dependence is
$t^{-0.6}$.

We have also obtained values for fit parameters ($\eps_e$, $\eps_B$, $p$ and $A_*$), keeping other parameters same, for GRB 190114C using the analytic approximation of the SSC emission, in order to compare with the fit parameters obtained from the smooth approximation. The values of the fit parameters from analytic approximation are listed in Table 1 within parenthesis. We note that while the values of $p$ and $\epsilon_e$ are comparable for both the approximations, the values of $A_*$ is a factor of three larger and $\epsilon_B$ is a factor of three smaller for the analytic approximation. This has implications on the $Y$-parameter and subsequently on determining the Thomson or Klein-Nishina regime for SSC emission. In case of the analytic approximation, the $Y$-parameter will be larger than the more accurate smooth approximation.

\section{Discussion}
In our modelling the blastwave is considered to be adiabatic and for GRB 090510 modelling we consider constant density ISM while for other two cases the medium is taken to be wind medium. This selection was inspired from the progenitor point of view for short and long GRBs, as most probably short GRBs bursts in a constant density medum while long GRBs takes place in the wind of progenitor star. The value of $\Gamma$ is used to tune the value of deceleration time $t_{\rm dec}$ such that the emission hits the afterglow phase. The fast to slow cooling transition time $t_{0}^{\rm ssc}$ in all three scenarios is such that the afterglow emission is explained using slow cooling.

We need $\Gamma = 1500$ to optimize the deceleration time for GRB 090510. In earlier works, value of $\Gamma =2000$ is used by \cite{2010A&A510L7G} and combination of larger $10^4$ and smaller value $220$ is used by \citep{corsi_two_bw} for narrow and wider jets. The low density medium used for GRB 090510 is also found in earlier work by \citep{corsi_two_bw} but higher value $0.1~ {\rm cm^{-3}}$ is also reported \citep{2016ApJ83122F}. The jet-break feature is found at $\sim$ 3 ks as indicated by the optical and X-ray light curves modelling and this is consistent with earlier work \citep{razzaque_aftrg_prot_sync}.

For long GRB 130427A and GRB 190114C in the optical emission for time larger than $3 \times 10^5$~s, we have added the host galaxy emission to the SEDs. In GRB 190114C the reason for the harder optical emission in between $10^4 - 10^5$ s could be refreshed shock \citep{1998ApJ496L1R, 2003Natur138G}. We found 10 times stronger wind in our model for GRB 130427A, $\dot{M}_{-5}/v_8 = 0.01$ compared to \cite{grb_model_1}, where they reported very weak stellar wind value 0.004. For GRB 190114C wind is 2 times stronger compared to GRB 130427A, i.e. $\dot{M}_{-5}/v_8 = 0.02$, which is lower compared to stellar wind value reported by the MAGIC paper \citep{2019Natu575459M}. In all three GRBs, we found that the internal $\gamma\gamma$ absorption in the blastwave is not important for the model parameters we have used but the redshift-dependent absorption is included to very high energy photons. In principle, gamma-rays absorbed in the EBL can initiate a cascade and secondary contributions to the overall emission can be important if the intergalactic magnetic field is $\lesssim 10^{-19}$~G \citep[see, e.g.,][]{2004ApJ...613.1072R, 2004MNRAS.354..414A, 2007ApJ...671.1886M}. We do not, however, discuss this here. The MAGIC discovery paper on GRB 190114C have discussed the importance of KN effects and attenuations in the jet will shape the emission \citep{2019Natu575459M}.  \citet{2019ApJ...880L..27D} found that Thomson scattering can be used in compared to the Klein-Nishina effect for this burst. We also found that the SSC emission in the Thomson regime can be used to model the sub-TeV observations of GRB~190114C. The $Y$ parameter in equation (13) is of the order of one for all three GRBs we have modelled. We have used the analytical and smooth approximation for the estimation of SSC flux in this work. These two scenarios are compared in Figure \ref{grb_190114_wind} for GRB 190114C where grey solid lines represent the analytical approximation and the dotted orange curves represent the smooth approximation. Using the sub-TeV component we can find Thomson approximation useful for the multi-wavelength interpretation of GRB 190114C. In other two GRBs we study the emission in the Thomson approximation only in this work.

\section{Conclusion}
In conclusions, our synchrotron-SSC modelling proves to be useful in fitting multiwavelength afterglow data from long and short GRBs,
including VHE data. Modelling these data sheds light on the GRB blastwave models and physical parameters involved in radio to VHE gamma-ray emission.
The predicted SSC emission is less dominant in GRB 090510 and significant for GRB 130427A. The detection of this component in GRB 190114C is
important to study the shock energy distributions in electrons and magnetic fields, and the environment surrounding the GRBs. The frequent detection of
TeV component in GRB afterglows by upcoming CTA and the Large High Altitude Air Shower Observatory (LHASSO) \citep{2019sctabookC,
2019arXiv190502773B} will enrich GRB afterglow models.

\section{Acknowledgements}
We thank the anonymous referee for his insightful   comments and suggestions. We are thankful to X.-Y.~Wang and N. Fraija and S. B. Pandey for reading the manuscript and comments, R.~J.~Britto for providing
the python script for the optical depth calculation, R. Liu, V. Chand for helpful discussions and M. Arimoto for his comments. The research work of J.C.J. was supported by a GES fellowship at the University of Johannesburg, where most part of the work
is completed. S.R.\ acknowledges support from the National Research Foundation (South Africa) with Grant No.\ 111749 (CPRR).

\section*{Data Availability}
The data used in this article are available in the article and in its online supplementary material. The code used for the SSC calculations can be shared on reasonable request to the corresponding author.

\footnotesize{
\bibliography{ref_prop}
}

 \renewcommand{\theequation}{A-\arabic{equation}}    
  \setcounter{equation}{0}  

\onecolumn

\section*{Appendix A}
\subsection*{Synchrotron self-absorption frequency}
The synchrotron spectra in equations~(\ref{fast_spectrum}) and (\ref{slow_spectrum}) have the lowest frequency break at $\nu_a$, below which the synchrotron spectrum becomes harder by an index 2/3 due to synchrotron-self-absorption \citep{radiat_process_Ry_Li}. We describe here the derivation of self-absorption frequency $\nu_a$ for the blastwave in the circumburst medium. To calculate this we first define the self-absorption coefficient based on \citep{granot_absor_freq},
\begin{equation}
 \alpha^{\prime}_{\nu^{\prime}} = \frac{p+2}{8 \pi m_e \nu{^{\prime}}^2} \int_{\gamma_{\rm m}}^{\infty} P^{\prime}(\gamma_e^{\prime}) \frac{N_e(\gamma_e)}{\gamma_e} d\gamma_e\,.
 \label{eq:abs_cof}
\end{equation}
Here the electron distribution $N_e(\gamma_e)$ is independent of the fast- and slow-cooling.  From the unmodified electron distribution, defined previously we have
\begin{equation}
N_e({\gamma_e}) = n^{\prime} (p-1) \gamma_{m}^{\prime^{p-1}} \gamma_e^{-p} \,,
\label{eq:elatg}
\end{equation}
where $n^{\prime}$ is the density of the electrons in the jet frame and is related to the ambient density $n(R)$, with $n^{\prime} \approx 4 \Gamma_{\rm g} n(R)$ \citep{blandford_mckee_soln_76, granot_absor_freq}
For the electron Lorentz factor $\gamma_e^{\prime}$ we can calculate the emitted frequency $\nu^{\prime}_{sy}$ and emitted power  $P^{\prime}(\gamma_e^{\prime})$ as \citep{granot_absor_freq}
\begin{equation}
\nu^{\prime}_{sy} = \frac{3 q_e B^{\prime}(t)\gamma_e^{{\prime}^2}\sin\alpha } {4 \pi m_e c}
\end{equation}
and
\begin{equation}
 P^{\prime}(\gamma_e^{\prime}) = \frac{2^{5/3} \pi q_e^3 B^{\prime}(t)\rm sin \alpha}{ \Gamma(\frac{1}{3}) m_e c^2} \left(\frac{\nu^{\prime}}{\nu^{\prime}_{sy}} \right)^{1/3}\,,
 \label{eq:powr}
\end{equation}
respectively. Here, $\Gamma$ represents the Gamma function. Using equations~(\ref{eq:abs_cof}-\ref{eq:powr}) we can derive the expression for the self-absorption coefficient as
\begin{equation}
 \alpha^{\prime}_{\nu^{\prime}} = \frac{(p+2)(p-1)n^{\prime}}{8 \pi m_e \nu^{{\prime^{5/3}}}} \frac{2^{5/3}\pi q_e^3 B^{\prime}(t) \rm{sin}\alpha}{\Gamma(\frac{1}{3}) m_ec^2} \times \left[\frac{4 \pi m_e c}{3 q_e  B^{\prime}(t) \rm{sin}\alpha}\right]^{1/3} \gamma_{m}^{\prime^{p-1}} \int_{\gamma_{m}^{\prime}}^{\infty} \gamma_e^{-(p+5/3)} d\gamma_e \,.
\end{equation}
We have further simplified this expression using an average value of $\rm{sin}^{3/2}\alpha$, which is equal to $(\sqrt{\pi}/5) \Gamma(1/3) \Gamma(5/6)$, as
\begin{equation}
\alpha^{\prime}_{\nu^{\prime}} = 4.7 \times 10^{-10} \left[ \frac{(p+2)(p-1)}{(3p+2)\nu^{{\prime^{5/3}}}} \right] q_e^{8/3} m_e^{-5/3}
\times m_p^{1/3} \epsilon_B^{1/3} \Gamma_{\rm g}(t)^{5/3} n(R)^{4/3} \gamma_m^{-5/3} \,.
\end{equation}
We derive the absorption coefficients in the ISM and wind cases as
\begin{equation}
\alpha^{\prime}_{\nu^{\prime}} ({\rm ISM}) =  925.6\, \frac{(p+2)(p-1)}{(3p+2)\nu^{{\prime^{5/3}}}} \epsilon_B^{1/3} n_0^{4/3} \gamma_{m}^{\prime^{-5/3}} \Gamma_{\rm g}(t)^{5/3} \,
\end{equation}
and
\begin{equation}
\alpha^{\prime}_{\nu^{\prime}} ({\rm Wind}) = 2\times 10^{50} \frac{(p+2)(p-1)}{(3p+2)\nu^{{\prime^{5/3}}}}
 \times \epsilon_B^{1/3} A_{\star}^{4/3} R(t)^{-8/3} \gamma_{m}^{\prime^{-5/3}} \Gamma_{\rm g}(t)^{5/3} \,,
\end{equation}
respectively.


From the above expressions of $\alpha^{\prime}_{\nu^{\prime}}$, at the absorption frequency $\nu^{\prime} = \nu_a^{\prime}$, the condition that must be satisfied is $\alpha^{\prime}_{\nu^{\prime}} R(t)/\Gamma_{\rm g}(t) =1$. Further, following \citet{book_dermer_09} for the slow-cooling case $\gamma_{e}^{\prime} = \gamma_{m}^{\prime}$ and for the fast-cooling case $\gamma_{e}^{\prime} = \gamma_{c}^{\prime}$ provides the general expression for the synchrotron-self-absorption frequency as
\begin{equation}
{\nu^{\prime}_{a [s,f]}} ({\rm ISM}) = 4253 \frac{(p+2)(p-1)}{(3p+2)}
\times \left[ \epsilon_B^{1/3} n_0^{4/3} \gamma_{[m,c]}^{\prime^{-5/3}} \Gamma_{\rm g}(t)^{8/3} 
\frac{ c t}{1+z} \right]^{3/5}
\label{eq:inism_nupr}
\end{equation}
and
\begin{equation}
{\nu^{\prime}_{a [s,f]}} ({\rm Wind}) =  8.3 \times 10^{48} \frac{(p+2)(p-1)}{(3p+2)}
\times \left[\epsilon_B^{1/3} A_{\star}^{4/3} \gamma_{[m,c]}^{\prime^{-5/3}} 
\Gamma_{\rm g}(t)^{-8/3} \left( \frac{ct}{1+z} \right)^{-5/3}  \right]^{3/5} ,
\label{eq:inwind_nupr}
\end{equation}
respectively, for the ISM and wind medium. The subscripts $[s,f] \to [m,c]$ refer to the slow- and fast-cooling cases. The self-absorption frequency depends on the spectral index $p$ of the electrons for both the fast- and slow-cooling scenarios, due to the their dependence on the minimum Lorentz factor $\gamma_m^{\prime}$. We report numerical values of $\nu_a$ in the Appendix for different blastwave evolution scenarios.

\begin{table}
\caption{The closure relations between the temporal index $\alpha$ and spectral index $\beta$ in various afterglow models for synchrotron and
inverse Compton emission with flux distribution $F_{\nu} \propto t^{\alpha} \nu^{\beta}$.
\label{Table_sy}} 
\begin{tabular}{llll}
\hline\hline  
& $\beta$ & $\alpha $   & $\alpha(\beta)$   \\
\hline
\multicolumn{4}{c}{Synchrotron emission} \\
\hline
Adiabatic (ISM) & \multicolumn{2}{c}{slow cooling}   \\
\hline
$ \nu < \nu_{a,sy}$         &  $2$    &   ${1/2}$     & $\beta/ 4$  \\
$\nu_{a,sy}\le \nu \le \nu_{m,sy}$   &  ${1/ 3}$ &   ${1/ 2}$ & $3 \beta / 2$\\
$\nu_{m,sy} < \nu <\nu_{c,sy}$   &  ${-(p-1)/2} $    &$ {-3(p-1) / 4}  $ & $3 \beta / 2$ \\
$\nu \ge \nu_{c,sy}$      &  ${- p/ 2} $    &   $- {(3p-2)/ 4} $  & ${(3 \beta +1) / 2}$  \\
\hline
Adiabatic (ISM) &  \multicolumn{2}{c}{fast cooling}  \\
\hline
$\nu< \nu_{a,sy}$          &  $2$    &   $1$     & $\beta / 2$  \\
$\nu_{a,sy} \le \nu \le\nu_{c,sy}$  &  ${1 / 3}$ &   ${1/ 6}$ &   ${\beta / 2}$ \\
$\nu_{c,sy} < \nu <\nu_{m,sy}$  &  ${-1/ 2}$    &   ${-1 / 4}$  & ${\beta / 2}$  \\
$\nu \ge \nu_{m,sy}$       &  ${-p / 2} $    &   ${-(3p-2)/ 4}$  & ${(3 \beta + 1) / 2}$  \\
\hline
Adiabatic (wind) & \multicolumn{2}{c}{slow cooling}  \\
\hline

$\nu< \nu_{a,sy}$         &  $2$    &   $1$     & $\beta / 2$  \\
$\nu_{a,sy} \le \nu \le \nu_{m,sy}$   &  ${1 / 3}$ &   $0 $ &   ${(3 \beta -1)/ 2}$ \\
$\nu_{m,sy} < \nu <\nu_{c,sy}$   &  ${-(p-1)/ 2}$    &$ -{(3p-1)/ 4} $ & $ {(3 \beta -1)/ 2} $ \\
$\nu \ge \nu_{c,sy}$      &  ${- p/ 2} $    &   $- {(3p-2)/ 4} $  & ${(3 \beta + 1) / 2}$  \\
\hline
Adiabatic (wind) & \multicolumn{2}{c}{fast cooling}  \\

$\nu< \nu_{a,sy}$          &  $2$    &   $2$     & $\beta$  \\
$\nu_{a,sy} \le \nu \le \nu_{c,sy}$  &  ${1 /3}$ &   $-2 / 3$ &   ${-(\beta +1) / 2}$ \\
$\nu_{c,sy} < \nu <\nu_{m,sy}$  &  ${-1/ 2}$    &   ${-1 / 4}$  & ${-(\beta +1) /2}$  \\
$\nu \ge \nu_{m,sy}$       &  ${-p / 2} $    &   ${-(3p-2) /4} $  & ${(3 \beta +1) / 2}$ \\
\hline
\multicolumn{4}{c}{SSC emission} \\
\hline
Adiabatic(ISM) & \multicolumn{2}{c}{slow cooling}  \\
\hline
$\nu< \nu_{a,ssc}$               & $2$  & ${9 / 4}$ &   $ 9\beta/8$ \\
$\nu_{a,ssc} \le \nu \le\nu_{m,ssc}$     &   ${1 /3}$ & ${1}$&   ${3 \beta }$\\
$\nu_{m,ssc} < \nu <\nu_{c,ssc}$        & $ {-(p-1) / 2} $ & $ {-(9p-11)/ 8}$& $ {(9 \beta +1) / 4}   $ \\
$\nu \ge \nu_{c,ssc}$           & ${-p/ 2} $ & ${-(9p-10)/ 8} $&   ${(9 \beta + 5) / 4}$  \\
\hline
Adiabatic(ISM) &  \multicolumn{2}{c}{fast cooling}  \\
\hline
$\nu< \nu_{a,ssc}$               & $2$  & ${3/ 4}$&   $ 3 \beta /8 $    \\
$\nu_{a,ssc} \le \nu \le\nu_{c,ssc}$   &   ${1 / 3}$ & ${1 / 3}$&   ${  \beta}$ \\
$\nu_{c,ssc} < \nu <\nu_{m,ssc}$    & $ {-1 / 2}$ & ${1 / 8}$&   ${ -\beta/ 4}$ \\
$\nu \ge \nu_{m,ssc}$           & ${-p/ 2} $ & ${-(9p-10)/ 8} $ &   ${(9 \beta +5) / 4}$  \\
\hline
Adiabatic (wind) & \multicolumn{2}{c}{slow cooling}  \\
\hline
$\nu< \nu_{a,ssc}$                 & $2$  & $3/2$ &   ${3} \beta /4 $\\
$\nu_{a,ssc} \le \nu \le\nu_{m,ssc}$   &   ${1/ 3}$ & $-1 / 3$ &   ${- \beta }$ \\
$\nu_{m,ssc} < \nu <\nu_{c,ssc}$      & $ {-(p-1)/ 2} $& $ -p $ &$ 2 \beta -1 $ \\
$\nu \ge \nu_{c,ssc}$           & ${-p / 2}$ & ${ -p+1}$  &   $ 2 \beta +1$\\
\hline
Adiabatic (wind) & \multicolumn{2}{c}{fast cooling}  \\
\hline
$\nu< \nu_{a,ssc}$                & $2$  & ${11/6}$ &   $ 11\beta /12 $\\
$\nu_{a,ssc} \le \nu \le\nu_{c,ssc}$    &   ${1 / 3}$ & ${-5/ 3}$&   $-5\beta $\\
$\nu_{c,ssc} < \nu <\nu_{m,ssc}$        & ${-1/ 2}$ & $0$&   $\beta + 1/2$ \\
$\nu \ge \nu_{m,ssc}$             & ${-p/ 2}$ & $-p+1$ &   ${ 2 \beta +1 } $\\
\hline
\hline
\end{tabular}
\label{Tab:alpha-beta}
\end{table}

\renewcommand{\theequation}{B-\arabic{equation}}    
  \setcounter{equation}{0}  

\section*{Appendix B}
Below we give numerical expressions for the blastwave evolution parameters, synchrotron parameters and break frequencies, and SSC parameters and break frequencies. These values are described for the adiabatic blastwaves when they propagate in the constant density medium (ISM) or in a wind-type environment. Here $d_{28} = d_L/10^{28}$ cm and $t_2 = t/100$ s and $\nu_{\rm eV} = \nu/{\rm 1eV}$.

\subsection*{Adiabatic blastwave in the constant density medium}

\be
\Gamma_{\rm g}  = 155 \,(1+z)^{3/8} n_0^{-1/8} E_{55}^{1/8} t_2^{-3/8}
\label{bw_G_ad_i}
\ee

\be
R = 1.2 \times 10^{18}
(1+z)^{-1/4} n_0^{-1/4} E_{55}^{1/4} t_2^{1/4} ~{\rm cm}
\label{bw_R_ad_i}
\ee

\be
B^{\prime} = 19.0 \,(1+z)^{3/8} \eps_{B,  -1}^{1/2} n_0^{3/8} 
E_{55}^{1/8} t_2^{-3/8} ~{\rm G} 
\label{bw_B_ad_i}
\ee

\ba
\gamma_m^{\prime} &=& 2.8 \times 10^{4} \left(\frac{p-2}{p-1}\right) (1+z)^{3/8} \eps_{e,-1} n_0^{-1/8}E_{55}^{1/8} t_2^{-3/8}
\label{rl_min}
\ea

\ba
\gamma_c^{\prime} &=& 137.8\, (1+z)^{-1/8} \eps_{B,-1}^{-1} n_0^{-5/8} E_{55}^{-3/8} t_2^{1/8} (1+Y)^{-1}
\ea

\ba
\gamma_s^{\prime} = 8.5 \times 10^{6} (1+z)^{-3/16} \eps_{B,-1}^{-1/4} n_0^{-3/16} \phi_1^{-1/2} E_{55}^{-1/16} t_2^{3/16} (1+Y)^{-1/2}
\ea

\ba
 h\nu_{a, \rm{fast}} = 9.0 \times 10^{-2} \left[\frac{(p+2)(p-1)}{(3p+2)}\right]^{3/5} (1+z)^{-1/2} \epsilon_{B,-1}^{6/5} n_0^{11/10} E_{55}^{7/10} t_2^{-1/2} (1+Y)  ~{\rm eV}
\ea

\ba
h\nu_{a, \rm{slow}} = 4.4 \times 10^{-4} \frac{(p+2)^{3/5}(p-1)^{8/5}}{(3p+2)^{3/5}(p-2)} (1+z)^{-1} \epsilon_{B,-1}^{1/5} \epsilon_{e,-1}^{-1}  n_0^{3/5} E_{55}^{1/5}  ~{\rm eV}
\ea

\ba
h\nu_c =  1.0 \, (1+z)^{-1/2} \epsilon_{B,-1}^{-3/2} n_0^{-1} E_{55}^{-1/2} t_2^{-1/2} (1+Y)^{-2} ~{\rm eV}
\ea

\ba
h\nu_m = 41.3 \left(\frac{p-2}{p-1}\right)^2\, (1+z)^{1/2} \epsilon_{B,-1}^{1/2} \epsilon_{e,-1}^2 E_{55}^{1/2} t_2^{-3/2} ~{\rm keV}
\label{break_nu_ad_ism_sy}
\ea

\ba
 h\nu_s = 3.7 \, (1+z)^{-5/8} \phi_1^{-1} n_0^{-1/8} E_{55}^{1/8} t_2^{-3/8}(1+Y)^{-1} ~{\rm GeV} 
\ea

\noindent
The above set of frequencies builds-up the spectral energy distribution for synchrotron emission. For these set of frequencies we also calculate the time when they will appear in the spectrum. We list two most frequent time breaks, $t_c$ and $t_m$,

\be
t_c = 94.1 (1+z)^{-1} \epsilon_{B,-1}^{-3} n_0^{-2} E_{55}^{-1}(1+Y)^{-4} \nu_{c, \rm eV}^{-2}~\rm{s}
\ee

\be
t_m = 1.2 \times 10^{5} \left(\frac{p-2}{p-1}\right)^{4/3} (1+z)^{1/3} \epsilon_{B,-1}^{1/3} \epsilon_{e,-1}^{4/3} E_{55}^{1/3} \nu_{m, \rm eV}^{-2/3}~ \rm{s}
\ee

\noindent
The synchrotron transition time for the fast to slow cooling is calculated for the time when $\nu_m$ and $\nu_c$ coincides, i.e.  $ \nu_m(t_0) = \nu_c(t_0) $, The synchrotron and effective inverse-Compton cooling times are given by,

\be
t_{0} = 4.3 \times 10^6 \left( \frac{p-2}{p-1}\right)^2(1+z) \eps_{B,-1}^2 \eps_{e,-1}^2 n_0 E_{55} ~{\rm s}
\label{t0_ad_ism}
\ee

\be
t_{0}^{\rm ssc} = 4.3 \times 10^6 \left( \frac{p-2}{p-1}\right)^2(1+z) \eps_{B,-1}^2 \eps_{e,-1}^2 n_0 E_{55}  (1+Y)^2~{\rm s}
\label{t0_ad_ism1}
\ee

\noindent
Now we have listed the set of break frequencies for the SSC component. The SSC break frequencies are,

\ba
 h\nu_{a,ssc, \rm{fast}} = 3.5\, \left[\frac{(p+2)(p-1)}{3p+2}\right]^{3/5}  (1+z)^{-3/4} \epsilon_{B,-1}^{-4/5} n_0^{-3/20} E_{55}^{-1/20} t_2^{-1/4} (1+Y)^{-1}~{\rm keV}
\ea

\ba
 h\nu_{a, ssc, \rm{slow}} = 0.7\, \left[\frac{ (p+2)^{3/5} (p-1)^{-2/5} (p-2)}{(3p+2)^{3/5}}\right] (1+z)^{-1/4} \epsilon_{B,-1}^{1/5} \epsilon_{e,-1} n_0^{7/20} E_{55}^{9/20}t_2^{-3/4} ~{\rm MeV}
\ea

\ba
 h\nu_{c, ssc} = 0.04 \, (1+z)^{-3/4}  \eps_{B,-1}^{-7/2} n_0^{-9/4} E_{55}^{-5/4} t_2^{-1/4} (1+Y)^{-4} ~{\rm MeV}
\ea

\ba
 h\nu_{m, ssc} = 66.5 \, \left(\frac{p-2}{p-1}\right)^4  (1+z)^{5/4} \eps_{B,-1}^{1/2} \eps_{e,-1}^4 n_0^{-1/4}E_{55}^{3/4} t_2^{-9/4} ~{\rm TeV}
\ea



\noindent
The break times for minimum and cooling frequencies are defined as,

\be
t_{c, ssc} = 2.6 \times 10^{20} (1+z)^{-3} \epsilon_{B,-1}^{-14} n_0^{-9} E_{55}^{-5} (1+Y)^{-16} \nu_{c, ssc,\rm eV}^{-4} ~\rm{s}
\ee
\be
t_{m, ssc} = 1.4 \times 10^{8} \left(\frac{p-2}{p-1}\right)^{16/9} (1+z)^{5/9} \epsilon_{B,-1}^{2/9} \epsilon_{e,-1}^{16/9} n_0^{-1/9} E_{55}^{1/3}
\nu_{m, ssc, \rm eV}^{-4/9}~\rm{s}
\ee
\noindent
The maximum flux values for the synchrotron and SSC emission are,
\ba
f_{\nu, \rm{max}} = 377.1\, (1+z)^{-1} \eps_{B,-1}^{1/2} n_0^{1/2} E_{55} d_{28}^{-2}~{\rm Jy}. 
\label{Fnumax_ad_ism_syn}
\ea
\ba
f_{\nu,{ \rm max}, ssc} = 7.7 \times 10^{-4} (1+z)^{-5/4} \eps_{B,-1}^{1/2} n_0^{5/4} E_{55}^{5/4} t_2^{1/4}d_{28}^{-2} ~{\rm Jy}.
\label{Fnumax_ad_ism_ic}
\ea

\subsection*{Adiabatic blastwave into the wind medium}
The parameters have the same physical meaning as for the expressions defined above.

\be
\Gamma_{\rm g}  = 113.6 \,(1+z)^{1/4} A_\star^{-1/4} E_{55}^{1/4} t_2^{-1/4}.
\label{bw_G_ad_w}
\ee

\be
R = 3.1 \times 10^{17} (1+z)^{-1/2} A_\star^{-1/2} E_{55}^{1/2} t_2^{1/2}~{\rm cm}. 
\label{bw_R_ad_w}
\ee
\be
B^\p =  24.8 \,(1+z)^{3/4} \eps_{B,-1}^{1/2} A_\star^{3/4} 
E_{55}^{-1/4} t_2^{-3/4} ~{\rm G}.
\label{bw_B_ad_w}
\ee

\be
\gamma_m^{\prime} = 2.1 \times 10^4 \left(\frac{p-2}{p-1}\right) (1+z)^{1/4} \eps_{e,-1} A_{\star}^{-1/4}E_{55}^{1/4} t_2^{-1/4}
\label{gm_w}
\ee

\be
\gamma_c^{\prime} = 111.0 (1+z)^{-3/4} \eps_{B,-1}^{-1} A_{\star}^{-5/4} E_{55}^{1/4} t_2^{3/4} (1+Y)^{-1}
\ee

\be
\gamma_s^{\prime} = 7.4 \times 10^{6} (1+z)^{-3/8} \eps_{B,-1}^{-1/4} \phi_1^{-1/2} A_{\star}^{-3/8} E_{55}^{1/8} t_2^{3/8} (1+Y)^{-1/2}
\ee
\be
h\nu_{a, \rm{fast}} = 4.3 \times 10^{-2} \left[\frac{(p-1)(p+2)}{(3p+2)} \right]^{3/5}  (1+z)^{3/5} \eps_{B,-1}^{6/5}  A_{\star}^{11/5} E_{55}^{-2/5} t_2^{-8/5} (1+Y) ~{\rm eV}.
\ee

\be
 h\nu_{a, \rm{slow}} = 2.3 \times 10^{-4} \frac{(p-1)^{8/5}(p+2)^{3/5}}{(3p+2)^{3/5}(p-2)} (1+z)^{-2/5} \eps_{B,-1}^{1/5} \eps_{e,-1}^{-1} A_{\star}^{6/5} E_{55}^{-2/5} t_2^{-3/5} ~{\rm eV}.
\ee

\ba
h\nu_c &=& 0.6 (1+z)^{-3/2} \eps_{B,-1}^{-3/2} A_\star^{-2} E_{55}^{1/2} t_2^{1/2} (1+Y)^{-2} ~{\rm eV}.
\ea

\ba
h\nu_m &=& 2.1 \times 10^{4}\, \left( \frac{p-2}{p-1} \right)^2(1+z)^{1/2} \eps_{B,-1}^{1/2} \eps_{e,-1}^2 E_{55}^{1/2} t_2^{-3/2} ~{\rm eV}.
\ea

\ba
h\nu_s &=& 2.7 \,  (1+z)^{-3/4} \phi_1^{-1} A_\star^{-1/4} E_{55}^{1/4} t_2^{-1/4} (1+Y)^{-1} ~{\rm GeV}.
\label{break_nu_ad_wind_sy}
\ea

\be
t_c = 1.6 \times 10^{2} (1+z)^{3} \epsilon_{B,-1}^{3} A_{\star}^4 E_{55}^{-1}(1+Y)^{4} \nu_{c, \rm eV}^{2}~\rm{s}
\ee

\be
t_m = 7.6 \times 10^{4} \left(\frac{p-2}{p-1}\right)^{4/3} (1+z)^{1/3} \epsilon_{B,-1}^{1/3} \epsilon_{e,-1}^{4/3} E_{55}^{1/3} \nu_{m, \rm eV}^{-2/3}~ \rm{s}
\ee

\be
t_0 = 2.0 \times 10^4 \left( \frac{p-2}{p-1}\right) (1+z) \eps_{B,-1} \eps_{e,-1} A_\star ~{\rm s} 
\label{t0_adiabatic_wind}
\ee
\be
t_{0}^{\rm ssc} = 2.0 \times 10^4 \left( \frac{p-2}{p-1}\right) (1+z) \eps_{B,-1} \eps_{e,-1} A_\star (1+Y) ~{\rm s}
\label{t0_adiabatic_wind1}
\ee

\ba
 h\nu_{a, ssc, \rm{fast}} = 1.1 \left[\frac{(p+2)(p-1)}{3p+2}\right]^{3/5}  (1+z)^{9/10} \epsilon_{B,-1}^{-4/5} A_{\star}^{-3/10}E_{55}^{1/10} t_2^{-1/10} (1+Y)^{-1} ~{\rm keV}.
\ea

\ba
 h\nu_{a, ssc, \rm{slow}} = 0.2\, \frac{(p-2)(p+2)^{3/5}}{(3p+2)^{3/5}(p-1)^{2/5}} (1+z)^{1/10} \epsilon_{B,-1}^{1/5} \epsilon_{e,-1} A_{\star}^{7/10}E_{55}^{1/10} t_2^{-11/10} ~{\rm MeV}.
\ea
\ba
h\nu_{c, ssc} &=& 15.0\, (1+z)^{-3} \eps_{B,-1}^{-7/2} A_\star^{-9/2} E_{55} t_2^{2} (1+Y)^{-4} ~{\rm keV}.
\ea

\ba
h\nu_{m, ssc} &=& 19.0\,\left( \frac{p-2}{p-1} \right)^4(1+z) \eps_{B,-1}^{1/2} \eps_{e,-1}^{4} E_{55} A_\star^{-1/2} t_2^{-2} ~{\rm TeV}.
\label{break_nu_ad_wind_ic}
\ea


\ba
t_{c, ssc} = 0.8 (1+z)^{3/2} \epsilon_{B,-1}^{7/4} A_{\star}^{9/4} E_{55}^{-1/2} (1+Y)^{2} \nu_{c, ssc, {\rm eV}}^{1/2} ~\rm{s}.
\ea
\ba
t_{m, ssc} = 4.3 \times 10^{8} \left(\frac{p-2}{p-1}\right)^{2} (1+z)^{1/2} \epsilon_{B,-1}^{1/4} \epsilon_{e,-1}^{2} E_{55}^{1/2} \nu_{m, ssc, {\rm eV}}^{-1/2}~\rm{s}
\ea

\ba
f_{\nu,{\rm max}} &=& 22.0\, (1+z)^{-1/2} \eps_{B,-1}^{1/2} A_\star E_{55}^{1/2} t_2^{-1/2} d_{28}^{-2}  ~{\rm Jy}.
\label{Fnumax_ad_wind}
\ea

\ba
f_{\nu, {\rm max}, ssc} =  3.8 \times 10^{-5} \eps_{B,-1}^{1/2} A_\star^{5/2} t_2^{-1} d_{28}^{-2}~{\rm Jy}.
\label{Fnumax_ad_ism}
\ea

\renewcommand{\theequation}{C-\arabic{equation}}    
  \setcounter{equation}{0}  

\end{document}